\documentclass[aps,preprintnumbers,showpacs,nofootinbib]{revtex4}
\usepackage{subfig}
\usepackage{graphicx}
\usepackage{dcolumn}
\usepackage{epsfig}
\usepackage{bm}
\newcommand{\eq}{\begin{eqnarray}}
\newcommand{\en}{\end{eqnarray}}

\newcommand{\ba}[1]{\begin{eqnarray} \label{(#1)}}
\newcommand{\ea}{\end{eqnarray}}

\newcommand{\newc}{\newcommand}
\newc{\lra}{\leftrightarrow}
\newc{\beq}{\begin{equation}}
\newc{\eeq}{\end{equation}}
\newc{\barr}{\begin{eqnarray}}
\newc{\earr}{\end{eqnarray}}
  \def\vbf{\mbox{\boldmath $\upsilon$}}
\begin{document}

\topmargin -0.50in
\title { Light WIMP searches involving electron scattering} 
\author{J.D. Vergados$^{1,2,4}$, Ch. C. Moustakidis$^3$, Yeuk-Kwan E. Cheung$^{4}$, H. Ejiri$^5$, Yeongduk Kim $^6$ and Jeong-Yeon Lee$^{6}$}
\affiliation{
$^1$TEI of Western Macedonia, Kozani, Gr 501 00,  Greece and  Center for Axion and Precision Physics Research, Institute
for Basic Science (IBS), Daejeon 34141, Republic of Korea\footnote{Permanent address, University of Ioannina, Ioannina, Gr 451 10, Greece.},}
\affiliation{$^2$CoEPP and Centre for the Subatomic Structure of Matter (CSSM), University of Adelaide, Adelaide SA 5005, Australia,}
\affiliation{$^3$ Department of Theoretical Physics, Aristotle University of
Thessaloniki, \\54124 Thessaloniki, Greece,}
\affiliation{4 Department of Physics, Nanjing University,
22 Hankou Road, Nanjing, China 210093,}
\affiliation{$^5$
 RCNP, Osaka University, Osaka, 567-0047, Japan}
\affiliation{$^6$
 Center for Underground Physics, IBS , Daejeon 34074, Republic of Korea.}
\begin{abstract}
We  consider light WIMP searches  involving the detection of recoiling electrons.
\end{abstract}
\pacs{ 93.35.+d 98.35.Gi 21.60.Cs}

\keywords{Dark matter, light WIMP,  direct  detection,  big bounce universe, WIMP-electron scattering, event rates, modulation}

\date{\today}
\begin{abstract}
In the present work we examine the possibility for detecting electrons in dark matter searches.  These detectors are considered to be the  most appropriate for detecting light dark matter particles with a mas in the MeV region. We analyze theoretically some key issues involved  in such a detection and we perform calculations for the expected rates employing reasonable theoretical models.
\end{abstract}
\maketitle
\section{Introduction}

The combined MAXIMA-1 \cite{MAXIMA1,MAXIMA2,MAXIMA3}, BOOMERANG \cite{BOOMERANG1,BOOMERANG2}
DASI \cite{DASI02} and COBE/DMR Cosmic Microwave Background (CMB)
observations \cite{COBE} imply that the Universe is flat
\cite{flat01}
and that most of the matter in
the universe is dark \cite{SPERGEL},  i.e. exotic. These results have been confirmed and improved
by the recent WMAP  \cite{WMAP06} and Planck \cite{PlanckCP13} data. Combining 
the data of these quite precise measurements one finds:
$$\Omega_b=0.0456 \pm 0.0015, \, \Omega _{\mbox{{\tiny CDM}}}=0.228 \pm 0.013 \mbox{ and } \Omega_{\Lambda}= 0.726 \pm 0.015~$$
(the more  recent Planck data yield a slightly different combination $ \Omega _{\mbox{{\tiny CDM}}}=0.274 \pm 0.020 , \quad \Omega_{\Lambda}= 0.686 \pm 0.020)$. It is worth mentioning that both the WMAP and the Plank observations yield essentially the same value of $\Omega_m h^2$,
  but they differ in the value of $h$, namely $h=0.704\pm0.013$ (WMAP) and $h=0.673\pm0.012$ (Planck).
Since any ``invisible" non exotic component cannot possibly exceed $40\%$ of the above $ \Omega _{\mbox{{\tiny CDM}}}$
~\cite {Benne}, exotic (non baryonic) matter is required and there is room for cold dark matter candidates or WIMPs (Weakly Interacting Massive Particles).\\
Even though there exists firm indirect evidence for a halo of dark matter
in galaxies from the
observed rotational curves, see e.g. the review \cite{UK01}, it is essential to directly
detect such matter in order to 
unravel the nature of the constituents of dark matter. 

The possibility of such detection, however, depends on the nature of the dark matter constituents and their interactions.

The WIMPs are  expected to have a velocity distribution with an average velocity which is close to the rotational velocity $\upsilon_0$ of the sun around the galaxy, i.e.  they are completely non relativistic. In fact a Maxwell-Boltzmann leads to a maximum energy transfer which is close to the average WIMP kinetic  energy  $\prec T\succ\approx 0.4\times 10^{-6}m c^2$. Thus for GeV WIMPS this average is in the keV regime, not high enough to excite the nucleus, but sufficient to measure the nuclear recoil energy. For light dark matter particles in the MeV region, which we will also  call WIMPs,  the average energy that can be transferred is  in the few eV region. So this light WIMPs  can be detected  by measuring the electron recoil after the collision. Electrons may of course be produced by heavy WIMPS after they collide with a heavy target which results in a shake up of the atom yielding "primordial" electron production \cite{VE04,EMouVer,MVE05}. This approach for sufficiently heavy WIMPs and target nuclei can produce electrons energies even in the  30 keV region, with a spectrum very different from that arising after a direct WIMP-electron collision. Furthermore WIMP-electron collisions involving WIMPs with masses in the few GeV region have also  recently appeared \cite{RFG16}-\cite{RCDFS16}. In the present work, however, we will restrict ourselves  in the case of light WIMPs with a mass in the region of the electron mass.

We will draw from the experience involving WIMPs in the GeV region. The event rate for such a process can
be computed from the following ingredients~\cite{LS96}: 
\begin{itemize}
\item [i)] The elementary electron cross section.
In this case we will consider the case of a scalar WIMP, whose mass, as far as we know has not been constrained by any experiment, but it leads to mass dependent cross section favoring light particles. This scalar WIMP couples with ordinary Higgs with a quartic coupling, the properties of which are being actively determined by the LHC experiments. Thus the WIMP interacts  with electrons via Higgs exchange with an amplitude proportional to the electron $f m_e$.
\item [ii)] The knowledge of the WIMP particle density in our vicinity.
 This is  extracted from WIMP density in the neighborhood of the solar system, obtained  from the rotation curves measurements. 
 The number  density of these  MeV WIMPs, however, is expected to be six orders of magnitude bigger than that of the standard WIMPs due to the smaller WIMP mass involved. 
\item [iii)] The WIMP velocity distribution.
In the present work we will consider a Maxwell-Boltzmann (MB) distribution.
\end{itemize}

In the electron recoil experiments, like the nuclear measurements first proposed more than 30 years ago \cite{GOODWIT}, one has to face the problem that the process of interest does not have a characteristic feature to distinguish it
from the background. So since low counting rates are  expected the background is
a formidable problem. Some special features of the WIMP- interaction can be exploited to reduce the background problems, such as  the modulation effect: This yields a periodic signal due to the motion of the earth around the sun. Unfortunately this effect, also proposed a long time ago~\cite{Druck} and subsequently studied by many authors~\cite{%
PSS88,GS93,RBERNABEI95,LS96,ABRIOLA98,HASENBALG98,JDV03,GREEN04,SFG06,FKLW11}, 
is small in the case of nuclear recoils, but we expect to be a bit larger in the case of the electron recoils. There has always been an interest in light WIMPs, see e.g. the recent work \cite{EMV12}. In fact the first direct detection limits on sub-GeV dark matter from XENON10 have recently been obtained \cite{EMMPV12}. This is encouraging, but based on our experience with standard nuclear recoil experiments to excited states \cite{VerEjSav13}, one has to make sure that the proper kinematics has to be used in dealing with bound electrons. Clearly  the binding electron energy plays  a similar role as the excitation energy of the nucleus, in determining the small fraction of the WIMP's energy to be transferred to the recoiling system.  It is therefore clear that 
Light WIMPs are quite different in  energy,  mass, interacting particle, and flux. 
Accordingly one needs detectors capable of  detecting low energy light WIMPs in the midst of formidable backgrounds, i.e. detectors  which are completely different from  current WIMP detectors employed for heavy WIMP searches. 

In the present paper  we will address the 
 implications of light scalar WIMPs on the expected event rates scattered off electrons.  The interest in such a WIMP has recently been revived due to a new scenario of dark matter production in bounce cosmology~\cite{Li:2014era, Cheung:2014nxi} in which the authors point out the possibility of using dark matter as a probe of a  big bounce at the early stage of cosmic evolution. 
A model independent study of dark matter production in the
contraction and expansion phases of the Big Bounce reveals a new venue for achieving the observed relic abundance in which dark matter was produced completely out of chemical equilibrium\cite{Cheung:2014pea} . 
In this way, this alternative route of dark matter production in bounce cosmology can be used to test the bounce cosmos hypothesis \cite{Cheung:2014pea}.
  
In any case, regardless of the validity of the big bounce universe scenario, the scalar WIMPs have the characteristic feature that the elementary cross section in their scattering off ordinary quarks or electrons  is increasing as the WIMPs get lighter, which leads to an interesting experimental feature, namely  it is expected to enhance the event rates at low WIMP mass. In the present calculation we will adopt  this view and study its implications in direct dark matter searches compared to other types of WIMPs, such as the neutralinos, which we will call standard WIMPs.

Scalar WIMP's can occur in particle models. Examples are i) In Kaluza-Klein theories for models involving    universal extra dimensions (for applications to direct dark matter detection  see, e.g.,~\cite{OikVerMou}). In such models  the scalar WIMPs are characterized by ordinary couplings, but they are expected to be quite massive, ii) extremely light  particles ~\cite{Fayet03}, which are not relevant to the ongoing WIMP searches, iii) Scalar WIMPS such  as those  considered previously in various extensions of the standard  model~\cite{Ma06}, which  can be quite  light and long lived protected by a discrete symmetry. Thus they are viable cold dark matter candidates.
\section{The particle model}
\begin{figure}[!ht]
\begin{center}
\subfloat[]
{
\includegraphics[width=0.6\textwidth]{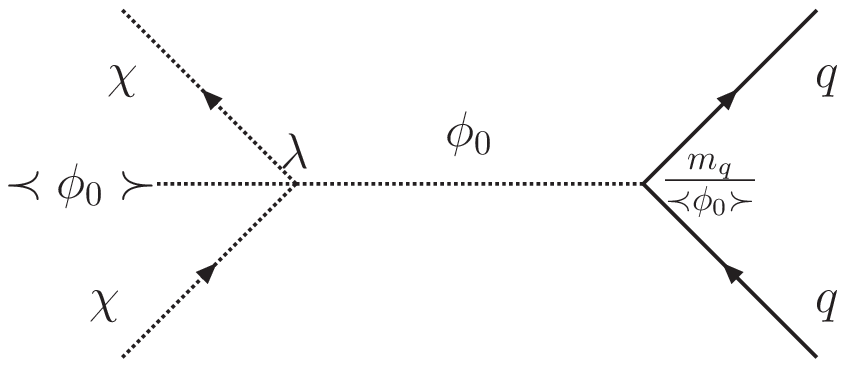}
}\\
\subfloat[]
{
\includegraphics[width=0.6\textwidth]{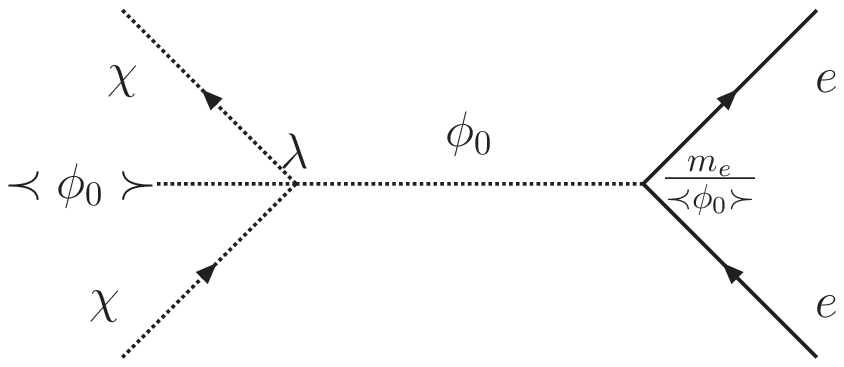}
}
\caption{(a) The quark - scalar WIMP  scattering mediated by a scalar particle. Note that the amplitude is independent of the vacuum expectation value $\prec\phi_0\succ$ of the scalar. (b) The corresponding diagram for electron-scalar WIMP
scattering.}
 \label{fig:xxphiphiqe}
 \end{center}
  \end{figure}

 The WIMP is assumed to be a scalar  particle $\chi$  interacting with another scalar $\phi$ via a quartic coupling  as discussed, e.g. in refs \cite{ZeeScal85,ZeeScal01,BentoRos01,BentoBero00}, and more recently  in Ref. \cite{Cheung:2014pea}.  
In fact the quartic coupling
 \beq
 \phi+\phi \rightarrow \chi+\chi
 \eeq 
involving the scalar WIMP $\chi$ and the Higgs $\phi$ leads to a Feynman diagram shown in  Fig.   \ref{fig:xxphiphiqe}(a) and results to a  mass dependent nucleon cross section \cite{Cheung:2014pea} of the form:
	  \beq
  \sigma_p=\sigma_0(p) \frac{1}{\left (1+m_{\chi}/m_p\right )^2},
	\eeq
	where  $\sigma_0(p)$ depends on the quartic coupling $\lambda$ and the quark structure of the nucleon. 
We will assume in this work that $\phi$ is the Higgs scalar discovered at LHC and the quartic coupling entering our model is the same with that involving the Higgs as has been  determined by the LHC experiments.

	  \begin{figure}[!ht]
\begin{center}
\subfloat[]
{
\rotatebox{90}{\hspace{0.0cm} $\sigma_e\rightarrow$pb}
\includegraphics[width=0.4\textwidth]{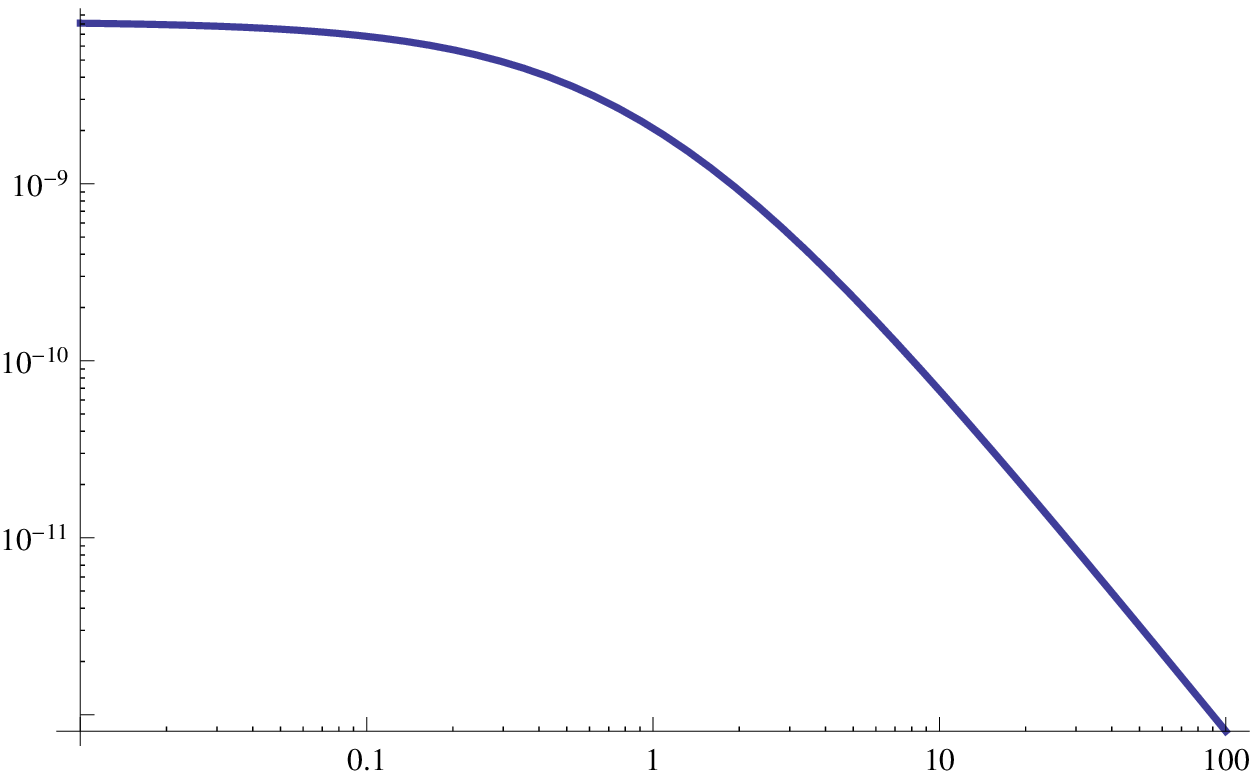}
}
\subfloat[]
{
\rotatebox{90}{\hspace{0.0cm} $\sigma_e\rightarrow$pb}
\includegraphics[width=0.4\textwidth]{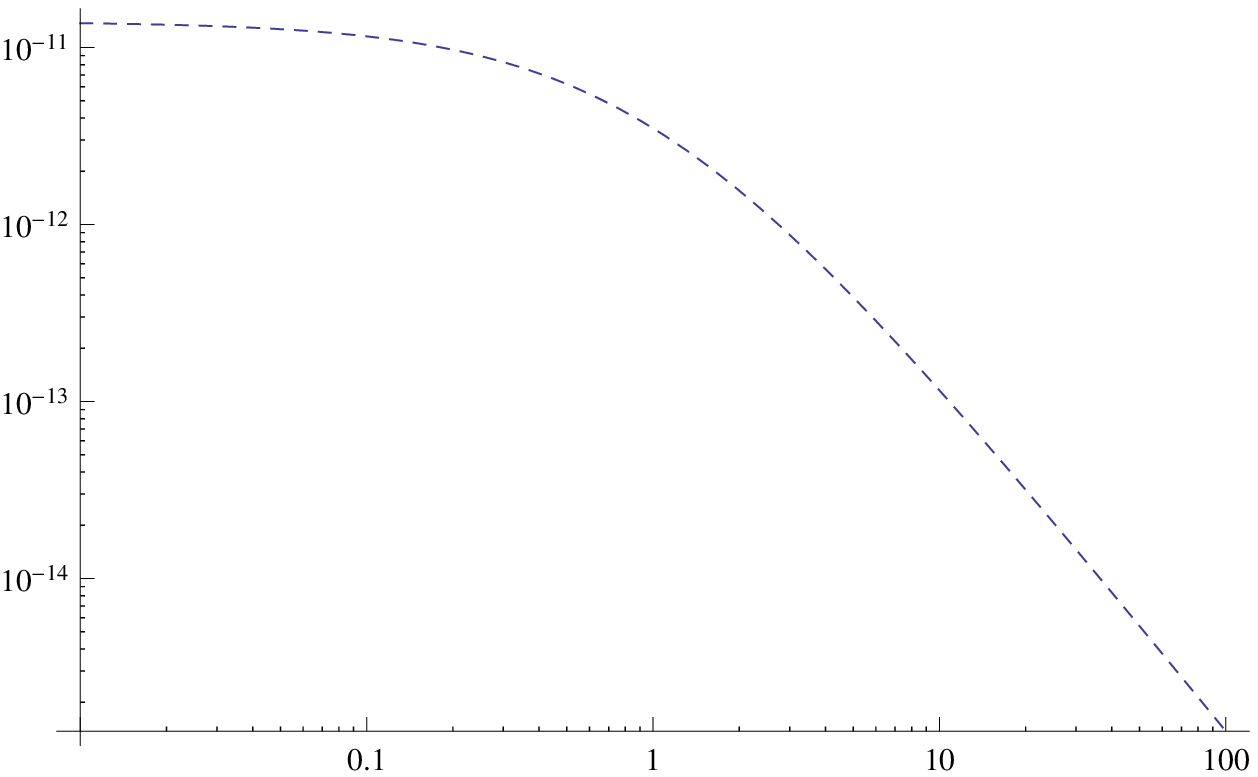}
}
\\{\hspace{-0.0cm}$\frac{m_{\chi}}{m_e} \rightarrow$ }
\caption{The WIMP-electron cross section in pb as a function of   WIMP mass, in units of the electron mass, as obtained our model in the case the WIMP is a scalar particle  with the value of $\lambda$ extracted from the Higgs LHC data (a) and that extracted from the nucleon cross section limit of $\sigma_p=10^{-8}$pb obtained from the exclusion plots of XENON100 at 50 GeV \cite{XENON10012} (b).}
 \label{fig:sigmape}
 \end{center}
  \end{figure}
	
  In the case of light WIMPs with mass less than 100 MeV  one cannot produce  a detectable recoiling nucleus, but  electrons~\cite{MVE05}  could be produced with energies in the  eV region, which, in principle,  could be detected with current mixed phase detectors ~\cite{XENON14}.  If the WIMP is a scalar particle,
	it can interact with electrons via the  Feynman diagram shown 
	in  Fig.~\ref{fig:xxphiphiqe}(b). 

	For WIMPs with mass in the range of  the electron mass, both the WIMP and the electron are not relativistic. So the expression for  elementary electron cross section is similar to that of hadrons , i.e. it is now given by:
	  \beq
  \sigma_e=\frac{1}{4 \pi}\frac{ \lambda^2 m_e^2}{m_{\phi}^4} \left (\frac{m_e m_{\chi}}{m_e+m_{\chi}}\right )^2\frac{1}{m_{\chi}^2}=\sigma_0(e) \frac{1}{\left (1+m_{\chi}/m_e\right )^2}
	\label{Eq:sigmae}
  \eeq
	with $\sigma_0(e) $ determined by the Higgs particle discovered at LHC, namely $\lambda=1/2$, $m_{\phi}=126$ GeV. One finds:
	\beq
	\sigma_0(e)=\frac{1}{4 \pi}\frac{ \lambda^2 m_e^2}{m_{\phi}^4}\approx 8.0 \times 10^{-9}\mbox{pb},
	\label{Eq:sigmae0}
	\eeq
	 This is a respectable size cross section, which depends on the ratio $m_{\chi}/m_e$.  Alternatively one could fix $\sigma_0 $ by using the nucleon cross limit~\cite{Cheung:2014pea} extracted from experiments, e.g. $\sigma_p=10^{-8}$ pb 
	 from  XENON100~\cite{XENON10012,XENON100.11}, 
	 which leads to $\sigma_0(p)=5.0 \times 10^{-5}$ pb. We thus find 
	\beq
	\sigma_0(e)=\sigma_0(p) \frac{m^2_e}{m_p^2}\approx 1.3\times 10^{-11} \mbox{ pb},
	\eeq
	which, if true, would  imply a much smaller value for $\lambda$.
	
	Anyway we will treat the value of $\sigma_0(e)$ as a parameter, which may be fixed if and when the experimental data become available.
	The  obtained cross section is exhibited in Fig. \ref{fig:sigmape}.  
	We note that this mass dependence of the cross section of scalar WIMPs results in a suppression of the cross section  in  the high WIMP mass regime. In what follows we will write $\sigma_0$ but it is understood that we mean  $\sigma_0(e)$.
	
	Before proceeding further with evaluation of the event rates in the case of light WIMPs it is instructive to review the experimental hurdles that must overcome to make their detection feasible.
	
\section{ Experimental aspects}
So far,  we have discussed mainly some theoretical aspects of light WIMPs in the MeV region and have presented  theoretical 
calculations of the expected  light WIMP signals. Here we briefly discuss experimental aspects of light WIMP searches, which are very different compared  
with those of heavy WIMP searches.
 
Light WIMPs are quite different in  energy,  mass, interacting particle, and flux. Accordingly 
one needs detectors which are completely different from  current WIMP detectors for heavy WIMPs. 
Detectors are required to observe 
low-energy light WIMP signals beyond/among  BG signals to identify the light WIMPs.  
Experimental aspects to be considered for light WIMP detectors are:
 i) the particle to be detected, 
 ii) the event rate,
iii) the  signal pulse height,
iv) the background rate and
v) the detector threshold energy. 
 We will now examine each of these items.
\begin{itemize}
\item[1] Particle to be detected.\\
Light WIMPs are detected by observing a recoil/scattered electron in the continuum region. 
In case that the WIMP interaction produces an ion-electron pair, one can detect the ion and/or the electron, and/or photons associated with the ion-electron pair.
If the recoil electron or the ion-electron pair energy is absorbed by the detector, one may measure the temperature change. 
These are similar to those from heavy WIMPs except that  their energies are very different. It is noted that atomic bound electrons are not excited by 
the light WIMPs  with $E\leq$ a few eV.  
\item[2] Event rate. \\
The cross section of $\sigma _0 \approx 8.2 \times 10^{-9}$ pb is an order of magnitude larger than the present XENON limit 
of $\sigma _0 \approx $ 10$^{-9}$ pb for heavy WIMPs~\cite{XENON10012}. 
The flux rate is around $n\approx 1.3 \times 10^{10} $cm$^{-2}$s$^{-1}$, 
which is larger by a factor of 50 GeV/0.5 MeV$\approx 10^5$ . 
Then the event rate for Xe detector is around $R\approx 1.8\times 10^{3}$ /(t.y), which is an order
 of magnitude larger than the present limit for 50 GeV heavy WIMPs \cite{XENON10012}.
\item[3] Signal pulse height.\\
The electron signal energy for light WIMPs is around 0.5-1 eV. This energy is 4 orders of magnitude smaller than
 the Xe nuclear recoil energy of around 25 keV for the 50 GeV WIMP.  
 The nuclear recoil signal is quenched by a factor 2-20, depending on the atomic number, in most heavy WIMP detectors.
 Thus the actual signal height for the light WIMP is 3 orders of magnitude 
smaller than that for the heavy WIMP. 
\item[4] Background rate\\
There are three types of background origins for WIMP detectors, radioactive (RI) 
impurities,  neutrons associated with cosmic rays, and electric noises.  
$\beta-\gamma$ rays from RI impurities produce BG electron signals, which are similar to electron signals from light WIMPs as well as 
 as those encountered in the case of double $\beta$ decay detectors, which measure $\beta $ rays. 
 BG rate for a typical future DBD (double $\beta$ decay ) detectors  is around 1/(t y keV) =10$^{-3}$/(t y eV) at a few MeV
 region \cite{ab14}. Then  one may expect a similar BG rate in the eV region. This is 3 orders of magnitude 
 smaller than the signal rate. Neutrons do not contribute to BGs in light WIMP detectors, 
although nuclear recoils from neutron nuclear reactions are most serious BGs for heavy WIMP detectors. 

Electric noises are most serious for light WIMP detectors because of the very low energy signals. 
The nuclear recoil energy from heavy WIMPs is typically a few 10 keV, and the signal pulse height is around 
a few keV if they are quenched, depending on the detector. This is of the same order of magnitude as electric noise levels.
Thus one can search for heavy WIMPs by measuring the higher velocity component above the electric noises 
On the other hand, the signal height for light WIMP is far below that of   typical  electric noises for current heavy WIMP detector.    
\item[5]  Energy threshold\\
The energy threshold $E_{th}$ for WIMP detectors is set necessarily below the WIMP signal, but just above the electric noise to be free from the noise. Then 
a very low energy threshold of an order of sub eV is required for light WIMP searches. This is 3-4 orders of magnitude smaller 
than the  level around 1-3 keV for most heavy WIMP detectors \cite{XENON10012} and \cite{ag13}. 
 Germanium semiconductor detectors are widely used to study low energy 
neutrinos and WIMPs. The ionization energy is 0.67 eV. Thus it can be used in principle for energetic light WIMPs. 
In practice, their threshold of around 200 eV \cite{ch14} or more is  still far above the light WIMP signals. Bolometers are, in principle, 
low energy threshold and high energy resolution detectors, but the energy threshold of  
practical 10 kg-scale bolometers are orders of magnitude higher than the light WIMP signal.  
Thus light WIMP detectors are necessarily different types from the present heavy WIMP detectors.

It is indeed a challenge to develop light WIMP detectors with low-threshold energy of the order of eV. Since the event rate is as large as 
2 10$^3$/t y, one can use a small volume detector of the order of 10 kgr at low temperature. In general, electric noises
are random in time. Then, coincidence measurements of two signals are quite effective to reduce electric noise signals in case that
 one light WIMP produces 2 or more  signals. One possible detector would be an ionization- scintillation detector, where one light WIMP interaction
produces one ionized ion and one electron. In case that the ionized ion traps an electron nearby and emits a scintillation photon, one may measure the primary
electron in coincidence with the scintillation photon.  Nuclear emulsion may be of potential interest for low energy electrons.  We briefly discuss possible new detection methods in section~\ref{sec:discussion}.
\end{itemize}
	
	\section{The differential WIMP-electron rate}
	\label{section:dWIMP-e-rate}
	The evaluation of the rate proceeds as in the case of the standard WIMP-nucleon scattering, but we will give the essential ingredients here to establish notation. We will begin by examining the case of a free electron.
	\subsection{Free electrons}
\begin{figure}[!ht]
  \begin{center}
\includegraphics[width=0.7\textwidth,height=0.4\textwidth]{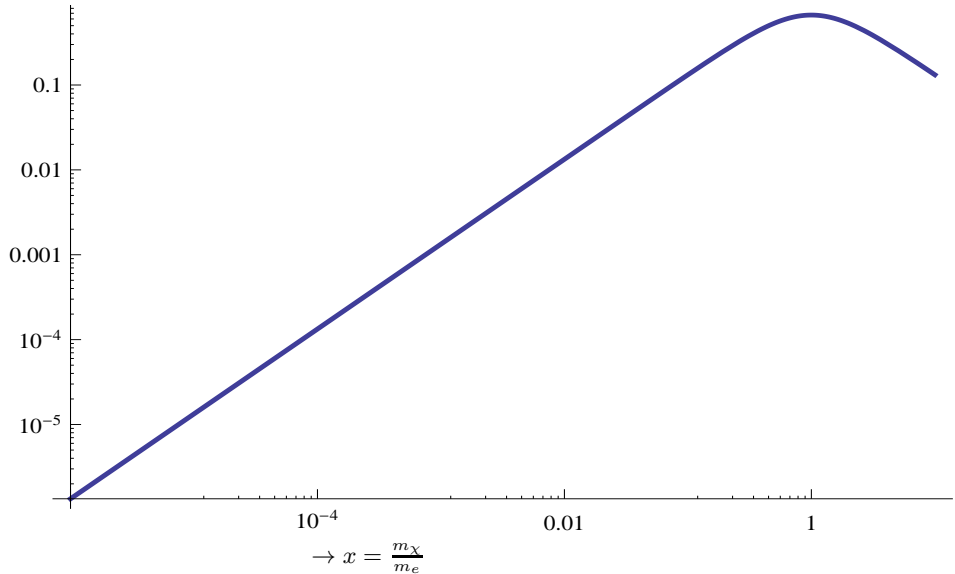}\\
\hspace{-3.0cm}$\rightarrow x=\frac{m_{\chi}}{m_e}$ \\
 \caption{The fraction of the WIMP energy transferred to the electron as a function of $x=\frac{m_{\chi}}{m_e}$}
 \label{fig:Txtometransf}
 \end{center}
  \end{figure}

	Since both the WIMP and the electron are not relativistic one finds that  the momentum transferred to the electron ${\bf q}$ has a magnitude is given by $$q=2 \mu_r \upsilon \xi,\,\mu_r=\frac{m_{\chi}m_e}{m_e+m_{\chi}}\mbox{=WIMP-electron reduced mass},\upsilon=\mbox{ WIMP velocity},\, \xi=\hat{\upsilon}.\hat{q}$$
	The differential cross section is now given by :
	\beq
	d\sigma=\sigma_0\frac{\mu_r^2}{m_{\chi}^2 }2\xi d\xi,\,0\le \xi\le 1
	\label{Eq.difsigma}
	\eeq
	From this, after integrating over $\xi$,  we obtain the total cross as given by Eq. \ref{Eq:sigmae}.\\ 
	The energy transfer is given by
	\beq
	 T=\frac{q^2}{2 m_e}=\frac{2 \mu_r^2 \upsilon^2 \xi^2}{m_e}
	\label{Eq:Ttransf}
	\eeq
	From this relation we find  that the fraction of the energy  of the WIMP  transferred to the electron, when taking $\left<\xi^2\right >=1/3$  for scattering at forward angles, is
	\beq
	\frac{T}{T_{\chi}}=\frac{4}{3} \frac{x}{(1+x)^2},\,x=\frac{m_e}{m_{\chi}}
	\label{Eq:Ttransfratio}
	\eeq
	This situation is exhibited in Fig. \ref{fig:Txtometransf}.
	We thus see that this fraction attains a   maximum when $x=1$, i.e., when the two masses are equal. Away from this value it becomes smaller. The effect is more crucial for very light WIMPs, since their average energy is much smaller.
	
	We also find from Eq. (\ref{Eq:Ttransf}) that the average energy of the electron 
	is given by
	\beq \left < T\right >=\frac{2}{3(1+x)^2} \left < \beta^2 \right > m_e c^2, \left < \beta^2 \right >=\prec\left (\frac{\upsilon}{c}\right )^2\succ=0.8 \times 10^{-6}
		\label{Eq:TtranseV}
	\eeq
Thus for MeV WIMPs the average energy transfer is in the eV region, which is reminiscent of the standard WIMPs where GeV mass leads to an energy transfer in the keV region. The maximum energy transfer corresponds to the escape velocity which is  $\upsilon_{esc}\approx 2 \sqrt{\left < \upsilon^2\right >}$, which leads to a value four times higher.  The exact expression of the maximum electron energy will be given below. 
	
It is preferable to rewrite Eq. \ref{Eq.difsigma} in terms of the energy of the recoiling electron T and the WIMP velocity using the relation given by Eq. (\ref{Eq:Ttransf}).
	We thus find
	\beq
	d\sigma=\sigma_0\frac{1}{2\upsilon^2} \frac{m_e}{m_{\chi^2}}dT
	\label{Eq.difsigma1}
	\eeq
	 We also find that 
	\beq
	\upsilon=\sqrt{\frac {m_eT}{2 \mu^2_r\xi^2}}\rightarrow \upsilon\ge\sqrt{\frac {m_eT}{2 \mu^2_r}}\rightarrow \upsilon_{min}=\sqrt{\frac {m_eT}{2 \mu^2_r}}
	\eeq
In other words the minimum velocity consistent with the energy transfer $T$ and the WIMP mass is constrained as above. The maximum velocity allowed is determined by the velocity distribution and it will be indicated by $\upsilon_{esc}$.
From this we can obtain the differential rate per electron in a given velocity volume $\upsilon^2 d \upsilon d \Omega$ as follows:
	\beq
	dR=\frac{\rho_{\chi}}{m_{\chi}}\upsilon \sigma_0\frac{1}{2} \frac{m_e}{m_{\chi^2}}dT f({\vbf}) d\upsilon d\Omega
	\eeq
	where $ f({\vbf})$ is the velocity distribution of WIMPs in the laboratory frame. Integrating over the allowed velocity distributions we obtain:
	\beq
	dR=\frac{\rho_{\chi}}{m_{\chi}} \sigma_0 \frac{1}{2}\frac{m_e}{m^2_{\chi}}dT \eta(\upsilon_{\mbox{\tiny{min}}}),\,\eta(\upsilon_{\mbox{\tiny{min}}})=\int_{\upsilon_{\mbox{\tiny{min}}}}^{\upsilon_{esc}} f({\vbf})\upsilon d\upsilon d\Omega
	\label{Eq:elrate1}
	\eeq
	 $\eta(\upsilon_{\mbox{\tiny{min}}})$ is a crucial parameter.\\
	Before proceeding further we find it convenient to express the velocities in units of the Sun's velocity. We should also take note of the fact the velocity distribution is given with respect to the center of the galaxy. For a M-B distribution this takes the form:
	\beq
	\frac{1}{\pi \sqrt{\pi}}e^{- y^{'2}},\,y^{'}=\frac{\upsilon^{'}}{\upsilon_0},\,\upsilon_0=220 \mbox{ km/s}
	\eeq
	We must transform it to the local coordinate system :
	\beq
	{\bf y}^{'}\rightarrow {\bf y}+{\hat\upsilon}_s+ \delta \left
(\sin{\alpha}{\hat x}-\cos{\alpha}\cos{\gamma}{\hat
y}+\cos{\alpha}\sin{\gamma} {\hat \upsilon}_s\right ) ,\,\delta=\frac{\upsilon_E}{\upsilon_0}
 \label{Eq:vlocal} \eeq 
with
$\gamma\approx \pi/6$, $ {\hat \upsilon}_s$ a unit vector in the
Sun's direction of motion, $\hat{x}$  a unit vector radially out
of the galaxy in our position and  $\hat{y}={\hat
\upsilon}_s\times \hat{x}$. The last term  in parenthesis in
 Eq. (\ref{Eq:vlocal}) corresponds to the motion of the Earth
around the Sun with $\upsilon_E\approx 28$ km/s being  the modulus of the
Earth's velocity around the Sun and $\alpha$ the phase of the Earth ($\alpha=0$ around June 3rd). The above formula assumes that the
motion  of both the Sun around the Galaxy and of the Earth around
the Sun are uniformly circular. Since $\delta$ is small, we can expand the distribution in powers of $\delta$ keeping terms  up to linear in $\delta$.
Then Eq. \ref{Eq:elrate1} can be cast in the form
\beq
	dR=\left (\frac{\rho_{\chi}}{m_{\chi}} {\upsilon_0}\right )\sigma_0 \frac{m_t Z_{eff}}{A m_p} 1.9\times  10^{6} \frac{1}{2}\frac{m_e}{m_{\chi^2}}  dT\left ( \Psi_0(y_{min})+
	 \Psi_1(y_{min}) \cos{\alpha} \right ),
	\label{Eq:elrate2}
	\eeq
where, in the above equation, the first term in parenthesis represents the average  flux of WIMPs, the second provides  the scale of the elementary cross section (in the present model the elementary cross section contains an additional mass dependence), the third term gives the number of electrons available for the scattering in a target of mass $m_t$ containing atoms with mass number $A$ and active electrons $Z_{eff}$ and the fourth is essentially the inverse of the square of the Sun's velocity in units of $c$ (its origin has its root in Eq. \ref{Eq.difsigma1}). Furthermore  for a M-B distribution one finds:
\beq
\Psi_0(x)=\frac{1}{2}H\left (y_{esc}-x \right )
  \left [\mbox{erf}(1-x)+\mbox{erf}(x+1)+\mbox{erfc}(1-y_{\mbox{\tiny{esc}}})+\mbox{erfc}(y_{\mbox{\tiny{esc}}}+1)-2 \right ]
  \label{Eq:Psi0MB}
\eeq
and
\barr
\Psi_1(x)&=&\frac{1}{2} H\left (y_{esc}-x \right )\delta 
   \left[\frac{ -\mbox{erf}(1-x)-\mbox{erf}(x+1)-\mbox{erfc}(1-y_{\mbox{\tiny{esc}}})-
   \mbox{erfc}(y_{\mbox{\tiny{esc}}}+1)}{2} \right . \nonumber\\
  && \left . +\frac{ e^{-(x-1)^2}}{\sqrt{\pi }}
   +\frac{
   e^{-(x+1)^2}}{\sqrt{\pi }}-\frac{ e^{-(y_{\mbox{\tiny{esc}}}-1)^2}}{\sqrt{\pi
   }}-\frac{ e^{-(y_{\mbox{\tiny{esc}}}+1)^2}}{\sqrt{\pi }}+1\right],
   \label{Eq:Psi1MB}
\earr
with
$$
y_{min}=\frac{\upsilon_{min}}{\upsilon_0}=\frac{1}{\upsilon_0}\sqrt{\frac {m_eT}{2 \mu^2_r}},\,y_{esc}=\frac{\upsilon_{esc}}{\upsilon_0}
$$
In the above expression  the Heaviside function $H$ guarantees that the required kinematical condition is satisfied. One can factor the constants out in the above equation to get
\beq
\frac{dR}{d(T/1\mbox{eV})}=\Lambda \left (\Sigma_0\left (\frac{m_{\chi}}{m_e},\frac{T}{1\mbox{eV}}\right )+\Sigma_1\left (\frac{m_{\chi}}{m_e},\frac{T}{1\mbox{eV}}\right ) \cos{\alpha} \right )
\label{Eq:ediffRate}
\eeq
where 
\beq
\Sigma_i(t,s)=\frac{1}{t^3}\Psi_i\left (1.23\left(1+\frac{1}{t}\right )\sqrt{s}\right ),\,i=0,1
\label{Eq:Sigma}
\eeq
and
\beq
	\Lambda=1.4 \frac{\rho_{\chi}}{m_e} {\upsilon_0}\sigma_0 \frac{m_t Z_{eff}}{A m_p} .
	\label{Eq:elrate3}
	\eeq
	The meaning of $Z_{eff} $ will become clear after we consider the fact that the electrons are not free but bound in the atom. Thus they are not all available for scattering, i.e.  $Z_{eff}<Z $ . 
	We will now estimate $\Lambda$ considering the following input:
	\begin{itemize}
	\item the elementary cross section $\sigma_0=8.2\times 10^{-9}pb=8.2 \times 10^{-45}\mbox{cm}^2$
	\item The total cross cross section, in units of $\sigma_0$, e.g.  $\sigma_{av}$ =0.2 for a WIMP mass about the electron mass (see below).
	\item The particle density of WIMPs in our vicinity:
	$$n =0.3 \times10^3 \mbox{(MeV /cm}^3\mbox{)/0.511MeV}\approx 600\mbox{cm}^{-3}$$
	(we use the electron mass in this estimate. The correct mass dependence has been included in evaluating $\sigma_{av}$). This value leads to a flux:
	$$ n\times 220 \mbox{ km/s}=1.3 \times 10^{10}\mbox{cm}^{-2}\mbox{s}^{-1}$$
	\item The number of electrons in a Kg of Xe:
	$$\left (1/(131 \times 1.67\times ^{-27})\right )Z_{eff}=4.6 \times 10^{24} Z_{eff}$$
	\end{itemize}
	Taking $Z_{eff}=54$, i.e. all electrons in Xe participating,  we expect about 
	$$\approx 0.5 \mbox{ events per kg.y}.$$
	 Encouraged by this estimate, even though it has been  obtained with a much smaller elementary cross section than previous estimates \cite{EMV12}, we are going to proceed in evaluating the expected spectrum of the recoiling electrons.
	
\begin{figure}[!ht]
  \begin{center}
\includegraphics[width=0.7\textwidth,height=0.5\textwidth]{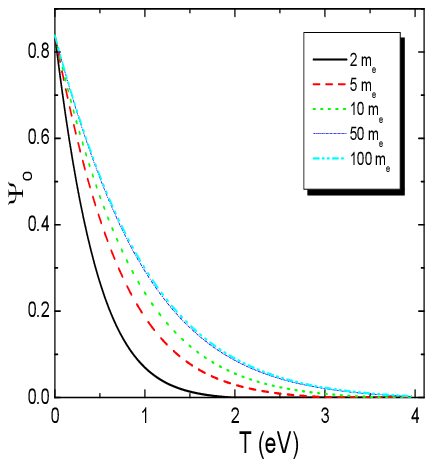}\\
 \caption{The shape of the differential cross section, as described by  the function $\Psi_0$, as a function of the electron recoil in eV for the WIMP masses $( 2, 5., 10, 50, 100)m_e$. The electrons are assumed to be free. As in the case of WIMPs the specrum does not exhibit any special structure. }
 \label{fig:Psi0}
 \end{center}
  \end{figure}

\begin{figure}[!ht]
  \begin{center}
\includegraphics[width=0.7\textwidth,height=0.6\textwidth]{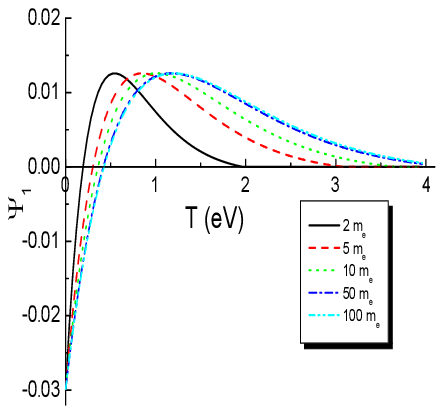}\\
 \caption{The same as in Fig. \ref{fig:Psi0} for the modulated amplitude described by the function $\Psi_1$,  as a function of  the electron energy $T$ in eV.}
 \label{fig:Psi1}
 \end{center}
  \end{figure}

\begin{figure}[!ht]
  \begin{center}
\includegraphics[width=0.7\textwidth,height=0.6\textwidth]{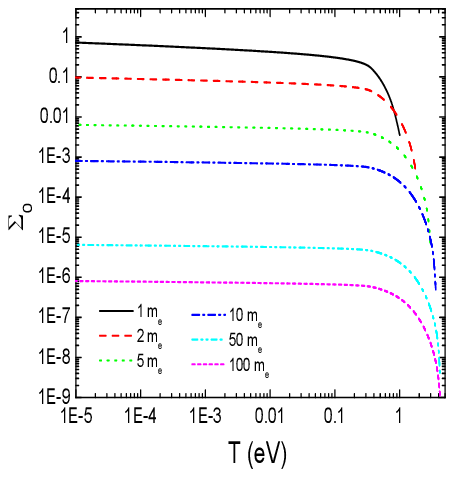}\\
 \caption{The shape of the differential cross section  as a function of the electron energy $T$ in eV for the WIMP masses $( 1,2, 5., 10, 50, 100)m_e$. The electrons are assumed to be free. The WIMP masses are indicated as in Fig. \ref{fig:Psi0}. The spectrum does not exhibit any special structure.}
 \label{fig:Sigma0}
 \end{center}
  \end{figure}

	We first show  the behavior of the the generic function $\Psi_0$ as a function of the electron energy for various WIMP masses (see Fig. \ref{fig:Psi0}). 
	In the case of the modulation amplitude we get the picture of  Fig. \ref{fig:Psi1}. We observe that the pattern is analogous to that found in the case of nuclear recoils.
	The function $\Sigma_0(t,s) $ is exhibited in Fig. \ref{fig:Sigma0}.
	
\begin{figure}[!ht]
  \begin{center}
	\subfloat[]
	{
\includegraphics[width=0.35\textwidth,height=0.3\textwidth]{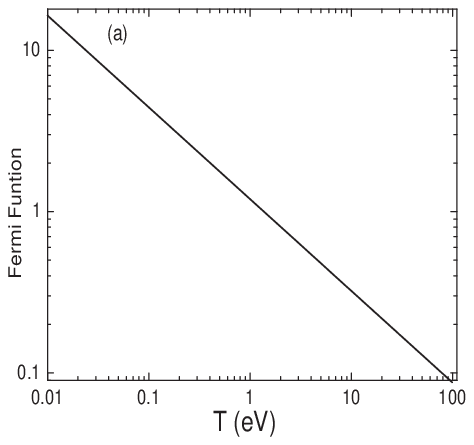}
}
\subfloat[]
	{
\includegraphics[width=0.35\textwidth,height=0.3\textwidth]{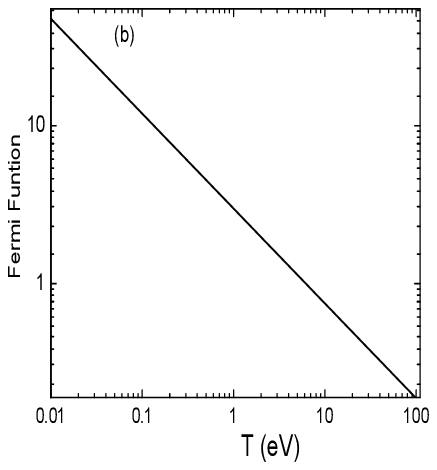}
}
 \caption{The Fermi function $F$ as a function of the electron recoil energy for $n=3$ and $Z=50$ (a) and $n=4$ and $Z=60$  (b) .}
 \label{fig:FermiFun}
 \end{center}
  \end{figure}

	 The various atomic physics approximations  involved involved involving relatively high electron, as e.g in recent works \cite{RFG16}-\cite{RCDFS16}, are not important in our case.The obtained rate, however, can increase substantially, if we include the correction of the outgoing electron wave due to the coulomb field. In beta decays this is done via the simple Fermi function \cite{VGBSS85}:
	\beq
	F(k,Z,\eta,\gamma)= \left ( k R\right )^{2 \gamma -2}e^{ \pi \nu}\left |\frac{\Gamma(\gamma+i \eta)}{\Gamma(2 \gamma )}\right |^2|M(\gamma+i\eta,2\gamma,2ikr)|^2,
	\eeq
	where 
	\beq
	\gamma=\sqrt{1-\alpha^2 Z^2},\,\eta=\alpha Z\frac{\sqrt{k^2 +m^2c^2}}{k} = Z\alpha \frac{T+m_e c^2}{\sqrt{T^2+2 m_e c^2 T}}
	\eeq
	and $M(\gamma+i\eta,2\gamma,2ikr)$ is  the Coulomb function represented by a confluent hypergeometric function. For $|(\gamma+i\eta)2ikr|<<2\gamma$, which is the case for momenta  and $r$ encountered in beta decay, the coulomb function becomes unity.  This is not true in our case. Furthermore  the first momentum dependent function,  employed  in the standard nuclear decay, is  different  in our case, since the electron is not produced at the origin, but it is ejected from an atomic orbit, i.e., $r$ is of the order  in Bohr orbit.
	The coulomb wave function
	for large  values of $\eta$ depends  on the variable $2 \eta k r=2 \alpha_Z m_e R   \approx 2 \alpha_Z m_e \bar{r}(n,Z,\gamma)$, where  $ \bar{r}(n,Z,\gamma)$ is the average radius, i.e it becomes  essentially independent  of the energy  $T$. It can be shown that \cite{LandauQm77} ${\bar r}(n,Z,\gamma)=\frac{1}{2}\left(3 n^2-(\gamma-1) \gamma)\right )\frac{1}{m_e\alpha Z}$. Thus $2\eta kr\approx\left(3 n^2-(\gamma-1)\gamma\right )$, i.e.
	\barr
	M(\gamma+i\eta,2\gamma,2ikr) &&\rightarrow\Gamma(2\gamma)f_c(\gamma,n),\nonumber\\
	f_c(\gamma,n)&\approx& \frac{2}{\sqrt{\pi}}\left (\sqrt{\left(3n^2-(\gamma-1) \gamma \right )}\right )^{-2\gamma+1/2} j_{2\gamma-1/2}\left (2\sqrt{\left(3n^2-(\gamma-1) \gamma \right )}\right )
		\earr
		where $j_{2\gamma-1/2}(y)$ is the spherical Bessel function. 
	Furthermore  to a good approximation :
	$$
	e^{ \pi \nu}|\Gamma(\gamma+i \eta)|^2\approx\frac{2 \pi  \eta}{1-e^{-2 \pi \eta}}
	$$
	(see, e.g., Landau's book  \cite{LandauQm77}). Thus one finds the Fermi function
	\beq
	F(T,Z,n,\eta,\gamma)\approx \left(\sqrt{T^2+2 m_eT} \bar{r}(n,Z,\gamma) \right )^{ 2 \gamma-2} f^2_c(\gamma,n)  \frac{2 \pi  \eta}{1-e^{-2 \pi \eta}}.
	\eeq
	 This function, which depends on the electron energy as well as the atomic parameters $n$ and $Z$ may lead to an enhancement for low energy electrons, see Fig. \ref{fig:FermiFun}. For a given $Z$, $F$ depends on $n$  and thus  a suitable average $n$ should perhaps be employed for a given target. Anyway the effect  of the Fermi function has not been included in Fig. \ref{fig:Sigma0}. 
	
\begin{figure}[!ht]
  \begin{center}
	\subfloat
	{
\includegraphics[width=0.7\textwidth,height=0.5\textwidth]{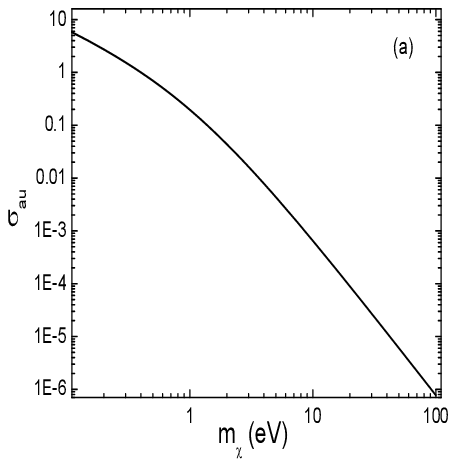}
}\\
\subfloat
	{
\includegraphics[width=0.4\textwidth,height=0.3\textwidth]{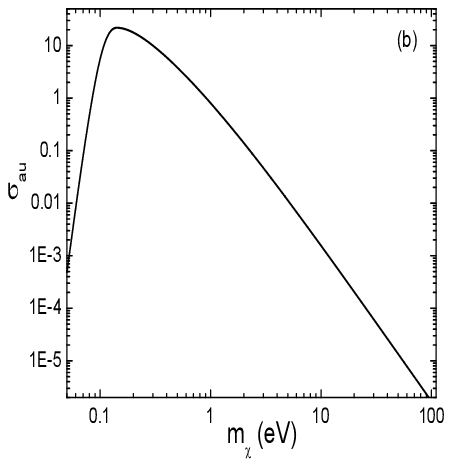}
}
\subfloat
	{
\includegraphics[width=0.4\textwidth,height=0.3\textwidth]{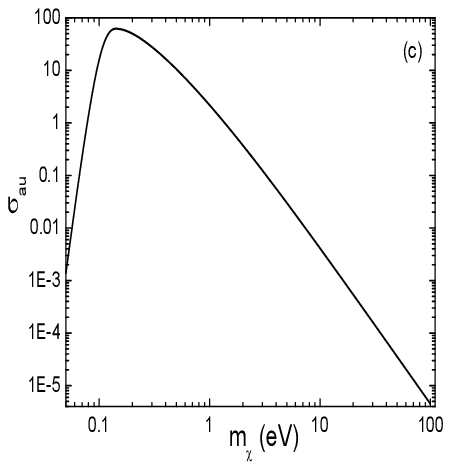}
}
 \caption{The total average WIMP-electron cross section in units $\sigma_0$ as a function of the WIMP mass in eV. In panel (a) the Fermi function was neglected, while in panels (b)  and (c) the Fermi function $F$ for $n=3$ and $Z=50$  and $n=4$ and $Z=60$ respectively has been employed.}
 \label{fig:totalav}
 \end{center}
  \end{figure}

	Integrating the differential rate given by Eq. (\ref{Eq:ediffRate}) over the electron spectrum we obtain the total rate:
	\beq
R=\Lambda \left ( \sigma_{av} +\sigma_{td}\cos{\alpha} \right )
\label{Eq:etotRate}
\eeq
where $\sigma_{av}$ and   $\sigma_{td}\cos{\alpha}$ are the average  and time dependent (modulated) cross sections respectively in units of $\sigma_0$, i.e. of the elementary cross section. The obtained quantity $\sigma_{av}$ is shown in Fig. \ref{fig:totalav} for free electrons both  without the Fermi function as well as with the Fermi function for two values $n$ and $Z$,  $n$ assumed to be sort of average. We see that the effect on the total cross section is large. Thus we find that the cross sections (in units of $\sigma_0$, for a  WIMP with the mass of the electron  we get values 0.2, 0.8 and 2.2
for cases (a), (b) and (c) respectively (see Fig. \ref{fig:totalav}), while for $n=2$ and $Z=50$ we find 0.3. In our estimates we will adopt an average enhancement factor   of 8 due to the Fermi function.
	Taking into account the Fermi function  the above estimate becomes 4.0 events/(kg.y).
	\section{Effects of binding of electrons}
	The binding of the electrons comes in in two ways. The first the most obvious. A portion of the energy of the WIMP will not go to recoil, but it will be  spent to release the bound electron. The second comes from the fact that the initial electron is not at rest but it has a momentum distribution, which is the Fourier  transform of its wave function in coordinate space.  We will first concentrate on the effect on the kinematics of the  binding energy.
	\subsection{The effect of the momentum distribution}
	section{The effect of the momentum distribution}
	In this case  of bound electrons the phase space reads:
	
	\beq
	J=\frac{1}{(2 \pi)^2}\int d^3{\bf p}\left |\tilde{\phi}_{n\ell}({\bf p} )\right |^2 d^3{\bf q}d^3{\bf p}_{\chi}^{'}\delta \left ({\bf p}+{\bf p}_{\chi}-{\bf p}_{\chi}^{'}-{\bf q}\right )\delta \left (\frac{{\bf p}^2_{\chi}}{2 m_{\chi}}-\frac{{\bf p}^{'2}_{\chi}}{2 m_{\chi}}-\frac{{\bf q}^2}{2 m_e^2}-b\right )
	\eeq
	where $ \tilde{\phi}_{n\ell}(p)$ is the electron wf in momentum space. After the integration over the momentum ${\bf p}$ via the $\delta$ function we obtain:
	\beq
	J=\frac{1}{(2 \pi)^2}\int\left |\tilde{\phi}_{n\ell}({\bf p'_{\chi}}-{\bf p_{\chi}}+{\bf q})\right |^2d^3{\bf p}_{\chi}^{'}d^3{\bf q}
	\delta \left (\frac{{\bf p}^2_{\chi}}{2 m_{\chi}}-\frac{{\bf p}^{'2}_{\chi}}{2 m_{\chi}}-\frac{{\bf q}^2}{2 m_e^2}-b\right )
	\eeq
	The integration over the magnitude of ${\bf p'_{\chi}}$ can be done using the energy conserving $\delta$ function and we obtain
	\beq
	J=\frac{1}{(2 \pi)^2}\int d^3{\bf q}Q^2 d\Omega_{{\bf Q}} \left |\tilde{\phi}_{n\ell} \left ({\bf Q}-{\bf p_{\chi}}+{\bf q}\right )\right |^2\frac{m_{\chi}}{Q},\,{\bf Q}=\hat{e}\sqrt{\left (m_{\chi}\upsilon\right )^2-q^2 x-2m_{\chi}b}
	\eeq
	with x=$m_{\chi}/m_e$ and $ \hat{e}$ a unit vector in the direction of ${\bf p'_{\chi}}$. We thus  find the important constraint:
	\beq
	\upsilon> \upsilon_{\mbox{\tiny min}},\,\upsilon_{\mbox{\tiny min}}=\frac{\sqrt{q^2 x+2 m_{\chi}b}}{m_{\chi}} \mbox{ or }\upsilon_{\mbox{\tiny min}}=\sqrt{\frac{2}{ m_{\chi}}}\sqrt{T+b}
	\eeq
	This already sets a limit on  the range  of  the   variables $b$ and $T $ of interest to experiments.
	\subsection{Range of $b$ and $T$}
\begin{figure}[!ht]
  \begin{center}
\rotatebox{90}{\hspace{-0.0cm} $b\rightarrow$ eV}
\includegraphics[width=0.7\textwidth,height=0.6\textwidth]{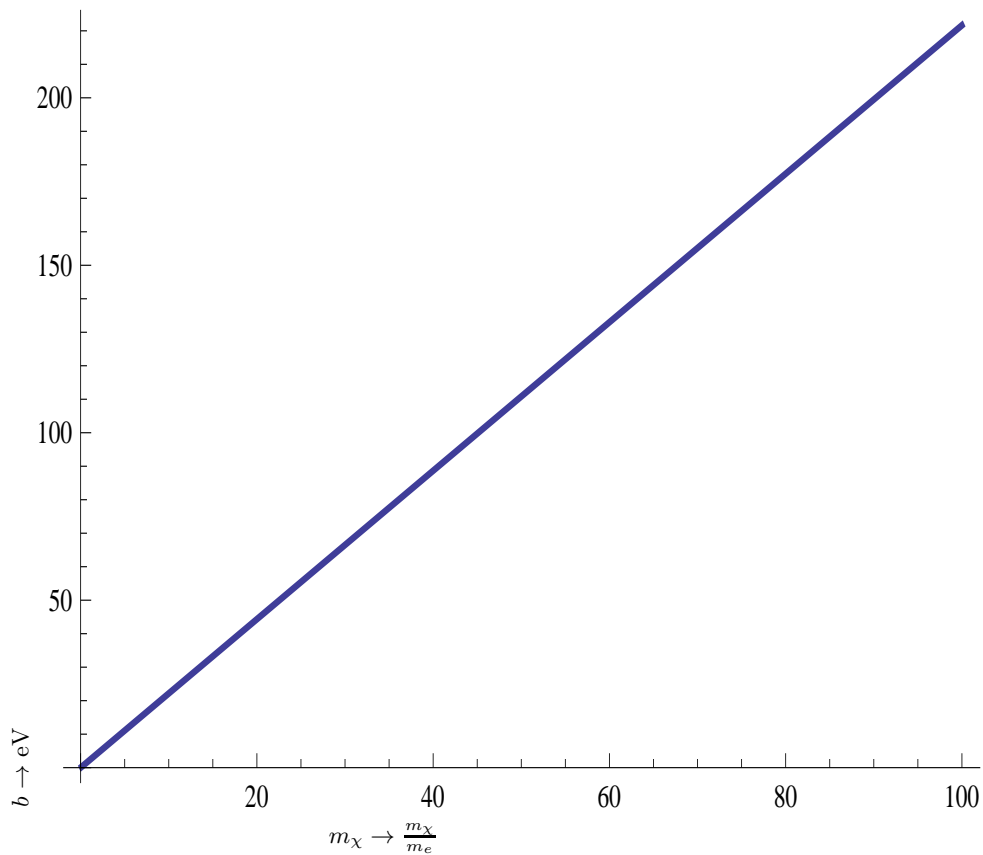}\\
\hspace{-3.0cm}$m_{\chi}\rightarrow \frac{m_{\chi}}{m_e}$ \\
 \caption{The  maximum binding electron energy $b_{ \mbox {\tiny{max}}}$ in eV as a function of $\frac{m_{\chi}}{m_e}$ accessible to WIMP-electron scattering.}
   \label{Fig.bvsmx}
 \end{center}
  \end{figure}

	Regarding $b$ the above conditions imply:
	$$(T)_{\mbox{\tiny{max}}}=\frac{1}{2}m_{\chi}\upsilon^2_{\mbox {\tiny{esc}}} -b,\, b_{\mbox{\tiny{max}}}=\frac{1}{2}m_{\chi}\upsilon^2_{\mbox {\tiny{esc}}}$$ 
	where  $b_{\mbox{\tiny{max}}}$ is associated with with zero maximum recoil energy and
	is exhibited as a function of WIMP mass in Fig. \ref{Fig.bvsmx}
	
\begin{figure}[!ht]
  \begin{center}
	\subfloat[]
	{
\rotatebox{90}{\hspace{-0.0cm} $\left ( \frac{m_{\chi}}{m_e}\right )_{\mbox{\tiny{min}}}\rightarrow$}
\includegraphics[width=0.7\textwidth,height=0.3\textwidth]{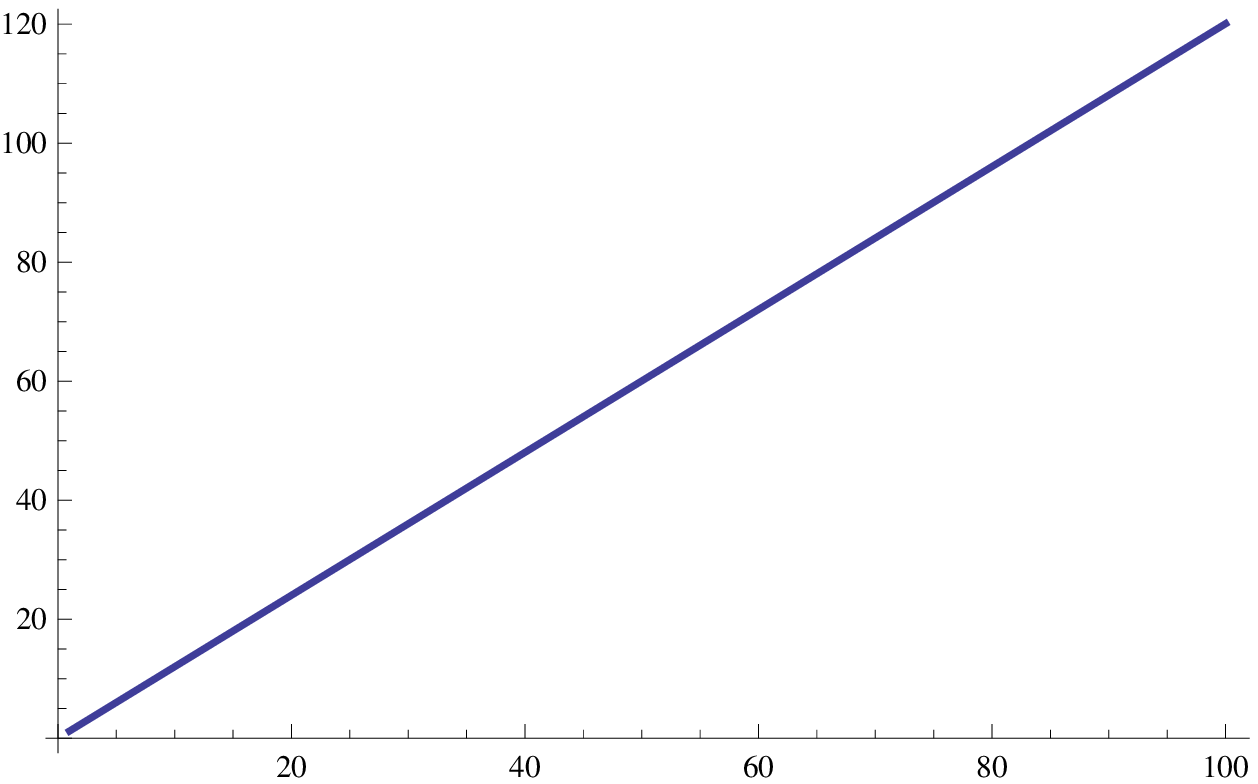}
}\\
\hspace{-3.0cm}$b\rightarrow$ eV \\
\subfloat[]
{
\rotatebox{90}{\hspace{-0.0cm} $T_{\mbox{\tiny{max}}}\rightarrow$ eV}
\includegraphics[width=0.7\textwidth,height=0.3\textwidth]{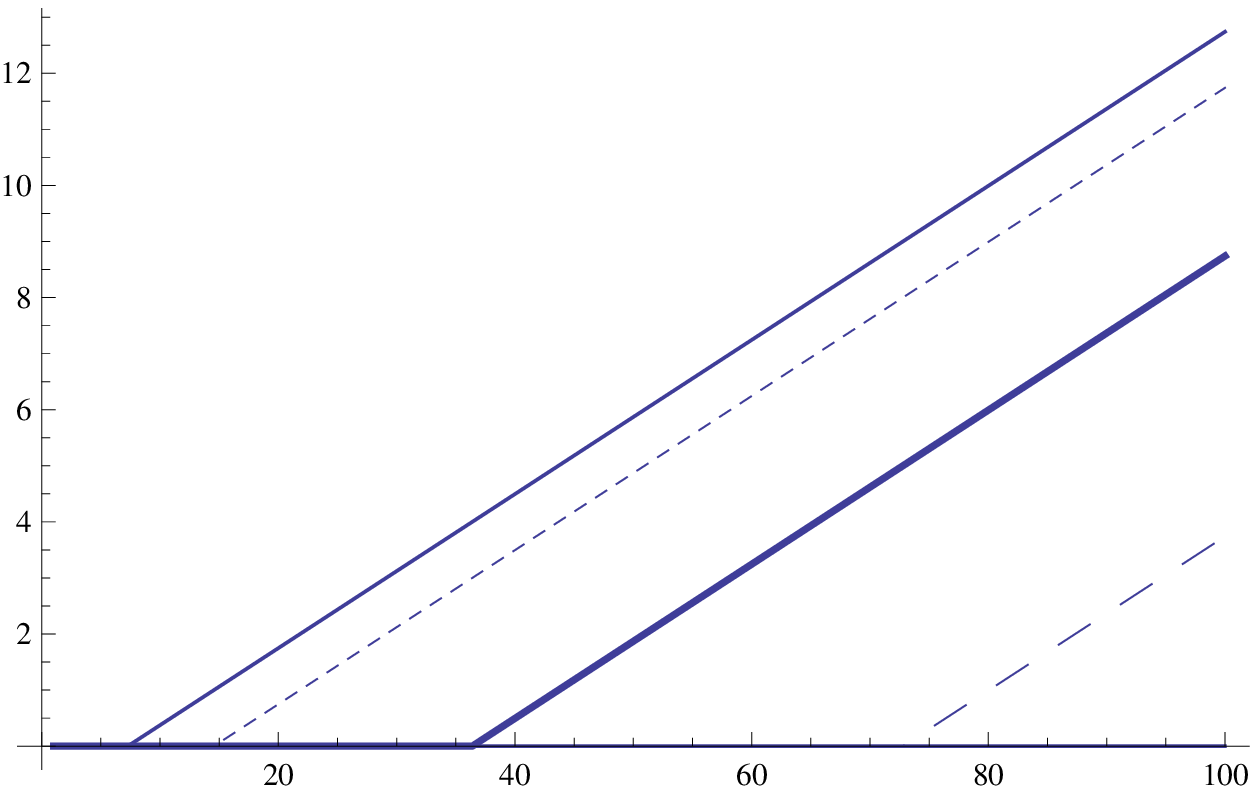}
}\\
\hspace{-3.0cm}$\rightarrow x=\frac{m_{\chi}}{m_e}$ 
 \caption{The  minimum $ \frac{m_{\chi}}{m_e}$ required to eject an electron with binding energy b in eV (a). The  maximum outgoing electron energy $ T_{\mbox{\tiny{max}}}$ as a function of $x=\frac{m_{\chi}}{m_e}$ for various binding energies  $b$ (b) . In this panel  the graphs correspond to $b=$1, 2, 5 and 10 eV  from left to right.}
   \label{Fig.bothxT}
 \end{center}
  \end{figure}

	What seriously affects detecting light WIMPs in the presence of large binding energies is the minimum WIMP velocity required to eject an electron. One finds that in order to surpass the barrier of a given binding energy $b$ the WIMP must have a minimum velocity at most $v_{\mbox{\tiny{esc}}}$ and a high mass, even if the electron energy is zero. This  the minimum $x=m_{\chi}/m_e$ required for this purpose is exhibited in Fig. \ref{Fig.bothxT}(a). The actual value of $x=m_{\chi}/m_e$ must, of course, be larger to get a reasonable rate.
	Finally we present in Fig. \ref{Fig.bothxT}(b) the maximum possible  energy for outgoing electrons as a function of the WIMP mass for various binding energies.
			\subsection{the phase space integral}
	To proceed further we must evaluate the above integral we must select a suitable coordinate system, e.g. one with the $z$-axis along the initial WIMP velocity,  the  $x$ axis in the direction of the outgoing WIMP and as $y$ axis perpendicular to the the plane of the other two. Then we find
	\barr
	J&=&\frac{1}{(2 \pi)}\int d^3{\bf q}m_{\chi}{Q}\int d\xi_1 d \phi \nonumber\\
	&&\left |\tilde{\phi}_{n\ell}\left (\sqrt{\left |Q^2+( m_{\chi}\upsilon )^2+q^2-2q m_{\chi}\upsilon \xi-2m_{\chi}\upsilon Q \xi_1+2 q Q\left (\xi \xi_1+\sqrt{1-\xi^2}\sqrt{1-\xi_1^2}\cos{\phi}
	\right)\right|}\right )Y^{\ell}_{m}(\hat{\omega})\right |^2
	\earr
	where $\hat{\omega}$ is a unit vector in the direction of ${\bf Q}-{\bf p_{\chi}}+{\bf q}$.\\
	There is no hope for obtaining an analytic expression with further approximations. We set $\cos{\phi}=0$ expecting that its contribution of this term  will average out to zero. The remaining integral is still complicated but for $s$ states we we find
	\beq
	J=\frac{1}{(2 \pi)}\int d^3{\bf q}m_{\chi}{Q}\int_{-1}^1 d\xi _1\left |\tilde{\phi}_{n\ell}\left (Q^2+( m_{\chi}\upsilon )^2+q^2
	-2 m_{\chi}q\upsilon \xi-2 m_{\chi} Q\upsilon \xi_1+2 q Q\xi \xi_1
	\right)\right |^2
	\eeq
	We now have to consider two cases:
	\begin{itemize}
	\item[i)] $q \xi\ne y$. Setting now 
	$$z=\frac{\sqrt{Q^2+( m_{\chi}\upsilon )^2+q^2
	-2 m_{\chi}q\upsilon \xi-2 m_{\chi} Q\upsilon \xi_1+2 q Q\xi \xi_1}}{p_0(nZ}
$$
where $p_0(n,Z)=(\alpha Z)/n) m_e$ the scale of momentum of the bound electron wave function  and $z$ dimensionless we obtain
\beq
J=\frac{1}{(2 \pi)} \int d^3{\bf q}m_{\chi}{Q}\frac{p^2_0(n,Z)}{Q( q \xi-m_{\chi}\upsilon)}\frac{1}{p_0^3(n,Z)}\int _{\eta_+}^{\eta_-} z dz \left |\tilde{\phi}_{n\ell}(z)\right |^2
\eeq
with $  \int z^2 dz \Omega \left |\tilde{\phi}_{n\ell}({\bf z})\right |^2$ normalized to one and
$$\eta_{\pm}=\frac{\sqrt{Q^2+q^2+(m_{\chi}\upsilon)^2-2 m_{\chi}q\upsilon \xi\pm 2Q(q\xi-m_{\chi} \upsilon) }}{p_0(n,Z)}$$
The integral over $z$ can be done analytically for $s$ states.
Thus
\beq
J=\frac{1}{(2 \pi)}\int d^3{\bf q}\frac{1}{p_0(n,Z)}\psi_1(\xi,q,\upsilon,b,p_0(n),\,\psi_1(\xi,q,\upsilon,b,p_0(n,Z))=\frac{m_{\chi}}{( q \xi-m_{\chi}\upsilon)}\int _{\eta_+}^{\eta_-} z dz \left |\tilde{\phi}_{n\ell}(z)\right |^2
\eeq
 Integrating over the angular part of ${\bf q}$ we obtain :
	\beq
	dJ= \frac{q}{p_0(n,Z)}q dq\psi_2(q,\upsilon,b,p_0(n,Z)),\,\psi_2(q,\upsilon,b,p_0(n,Z))=\int_{-1}^{1} d \xi \psi_1(\xi,q,\upsilon,b,p_0(n,Z))
	\eeq
	\item[i)] $q\xi=y$. Then
	\beq
	dJ= \frac{m_{\chi} Q}{y p_0(n,Z) }q dq\psi'_2(q,\upsilon,b,p_0(n,Z)),\,\psi'_2(q,\upsilon,b,p_0(n,Z))
	\eeq
	with
	\beq
	\psi'_2(q,\upsilon,b,p_0(n,Z))=\int_{(Q^2+(y-q)^2)/p^2_0(n,Z)}^{(Q^2+(y+q)^2)/p^2_0(n,Z)} z dz\left |\tilde{\phi}_{n\ell}(z)\right |^2
	\eeq
	This situation does not occur in practice.
	\end{itemize}
	The above lead to a cross section:
	\beq
	d\sigma=\frac{1}{\upsilon}|{\cal A}|^2 dJ
	\eeq
	where ${\cal A} $ is the invariant amplitude for the process. In the case of scalar WIMPs we have:
	\beq
	|{\cal A}|^2=\frac{(\lambda m_e)^2}{m_H^4}\frac{1}{(2 m_{\chi})^2},
	\eeq
	where the last factor comes from the normalization of the scalar field.
	By setting
	$$ \frac{(\lambda m_e)^2}{m_H^4}=4\pi\sigma_0$$ 
	we find 
	\beq
	d\sigma=\sigma_0 \frac{1}{\upsilon}\frac{\pi}{m_{\chi}^2} dJ
	\label{Eq:crosection}
	\eeq
	
	One can cast Eq. (\ref{Eq:crosection}) in a form similar to the expression for free electrons, namely
	\beq
	d\sigma=\sigma_0\frac{1}{2\upsilon^2} \frac{m_e}{m_{\chi^2}}dT
	\label{Eq.difsigma2}
	\eeq
	\beq
	d \sigma=\sigma_0\frac{1}{ \upsilon^2} \frac{m_e}{m_{\chi}^2} dT \tilde{\Lambda}(T,\upsilon,b,p_0(n,Z)),\,\tilde{\Lambda}(T,\upsilon,b,p_0(n,Z))=2\pi \upsilon\frac{\sqrt{2 m_e T}}{p_0(n,Z)}\psi_2(\sqrt{2 m_e T},\upsilon,b,p_0(n,Z))
	\eeq
  In order to get the differential rate one must fold the above expression with the velocity distribution $f(\vbf)$ 
	\beq
	\frac{d R}{dT}\propto \sigma_0 \frac{m_e}{m_{\chi}^2} \left[ \int_{\sqrt{2(b+T)/m_{\chi}}}^{\upsilon_{\mbox{\tiny{esc}}}}  \tilde{\Lambda}(T,\upsilon,b,p_0(n,Z))f(\vbf)\upsilon d \upsilon d \hat{\vbf} \right ]
	\eeq
	Note the appearance of the quantities $b$ and $\tilde{\Lambda}$ as a result of the electron binding. The behavior of the  function $\tilde{\Lambda}$  as a function of the velocity and its numerical value will significantly affect the obtained rates.  In order to obtain it the 
	 remaining integrals must be done numerically for each electron orbit separately , which is not trivial. This is currently under study, but at present will report results obtained in an approximate scheme valid for relatively low mass WIMPs, which is adequate for our purposes.
	\subsection{A convenient approximation for light WIMPs}
	As we have said in the case of scalar WIMPs the cross section is suppressed for WIMPs much heavier than the electron. In this case the momentum of the outgoing electron is small compared to the $p_0(n,Z)$. Let us assume that:
	$$\frac{Q (m_{\chi \upsilon })}{p^2_0(n,Z)}<1$$
	This means that 
	$$x^2 \upsilon \sqrt{\upsilon-\upsilon_1}\le  \alpha^2,\,\upsilon_1=\sqrt{2 (b+T)/x m_e} \le \upsilon_{\mbox{\tiny esc}}$$
	This   quantity as small so long as
	$$x<\frac{(\alpha Z)}{\upsilon_{\mbox{\tiny esc}}}=1.2 Z\approx 50\mbox { for a large atom  }$$ 
	For such values of $x$ we can expand the integral $ \int _{\eta_+}^{\eta_-} z dz \left |\tilde{\phi}_{n\ell}(z)\right |^2
$ up to second order in the small parameter. The result, e.g. for $1s$ hydrogenic wave functions, is:
$$J_1=\frac{16Q(m_{\chi}\upsilon-q \xi)}{\pi^2 p^2_0(n,Z)}$$
Integrating over the angles of the outgoing electron we obtain
\beq
dJ=q^2 dq \frac{64 Q^2m_{\chi} m_{\chi}\upsilon}{\pi p^5_0 (n,Z)}
\eeq
Proceeding as above we obtain
\beq
\tilde{\Lambda}=\tilde{\Lambda}_0(Z) \sqrt{\frac{2 T}{m_e}} \frac{x^4}{(\alpha Z)^5} \upsilon^2(\upsilon^2-\upsilon_1^2)
\eeq
or
\beq
\tilde{\Lambda}=\tilde{\Lambda}_0 (Z) x^4 \sqrt{T} y^2 \left(y^2-\frac{6.59 (b+T)}{x}\right),\, \tilde{\Lambda}_0(1)=3.5\times 10^{-3}
\eeq
where $b$ and $T$ are in eV and $y$ is the WIMP velocity in units of the sun's velocity. A similar expression with a slightly different  constant $\tilde{\Lambda}_0 (Z)$ is expected to hold for other electron orbitals.

\begin{figure}[!ht]
  \begin{center}
\rotatebox{90}{\hspace{-0.0cm} $f(\frac{\upsilon}{\upsilon_0})\rightarrow$}
\includegraphics[width=0.7\textwidth,height=0.5\textwidth]{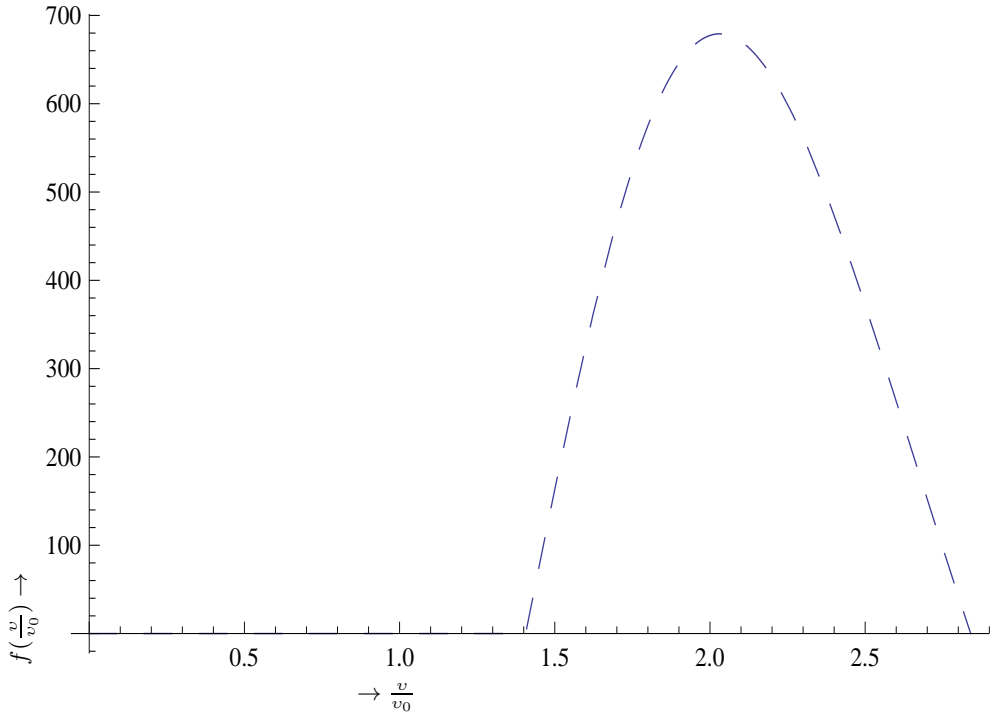}\\
\hspace{-3.0cm}$\rightarrow \frac{\upsilon}{\upsilon_0}$ \\
 \caption{The  velocity distribution obtaine with $x=2$  and $b=T=1$ eV. Note the lower bound on the velocity to the binding of the electrons.}
   \label{Fig.vbTx2}
 \end{center}
  \end{figure}

\begin{figure}[!ht]
  \begin{center}
\rotatebox{90}{\hspace{-0.0cm} $f(\frac{\upsilon}{\upsilon_0})\rightarrow$}
\includegraphics[width=0.7\textwidth,height=0.5\textwidth]{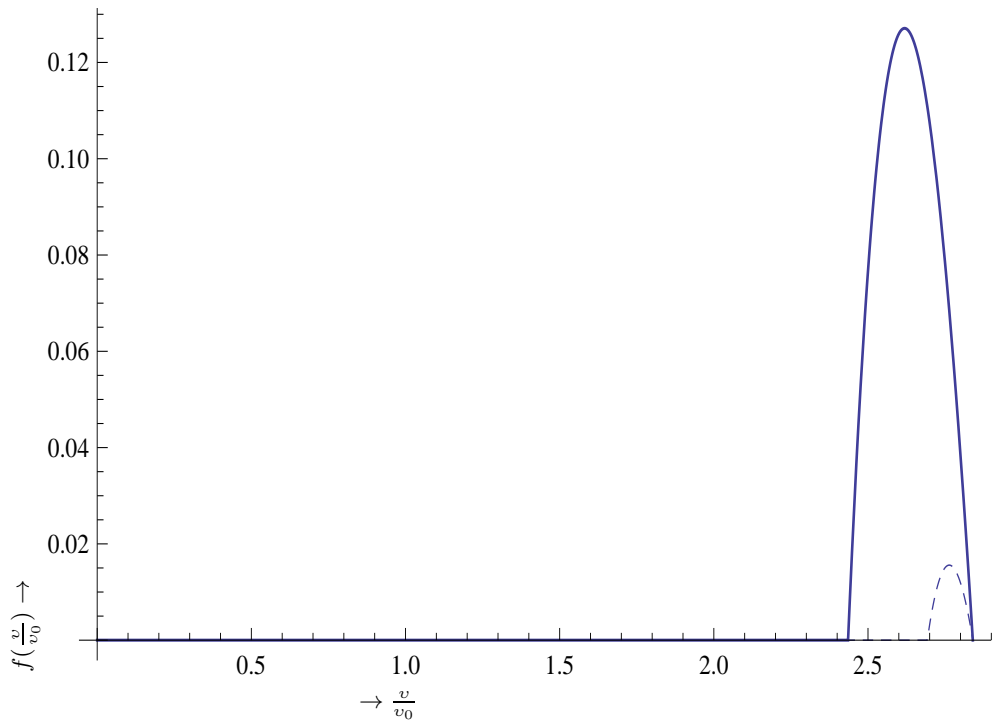}\\
\hspace{-3.0cm}$\rightarrow \frac{\upsilon}{\upsilon_0}$ \\
 \caption{The  velocity distribution obtained with $x=5$, $T=2$eV  and $b=2.5$ and 3.5 eV associated with the solid and dashed line respectively. Note the large effect of the binding both on the scale and the velocity range for so light WIMPs.}
   \label{Fig.vbTx5}
 \end{center}
  \end{figure}

\begin{figure}[!ht]
  \begin{center}
\rotatebox{90}{\hspace{-0.0cm} $f(\frac{\upsilon}{\upsilon_0})\rightarrow$}
\includegraphics[width=0.7\textwidth,height=0.5\textwidth]{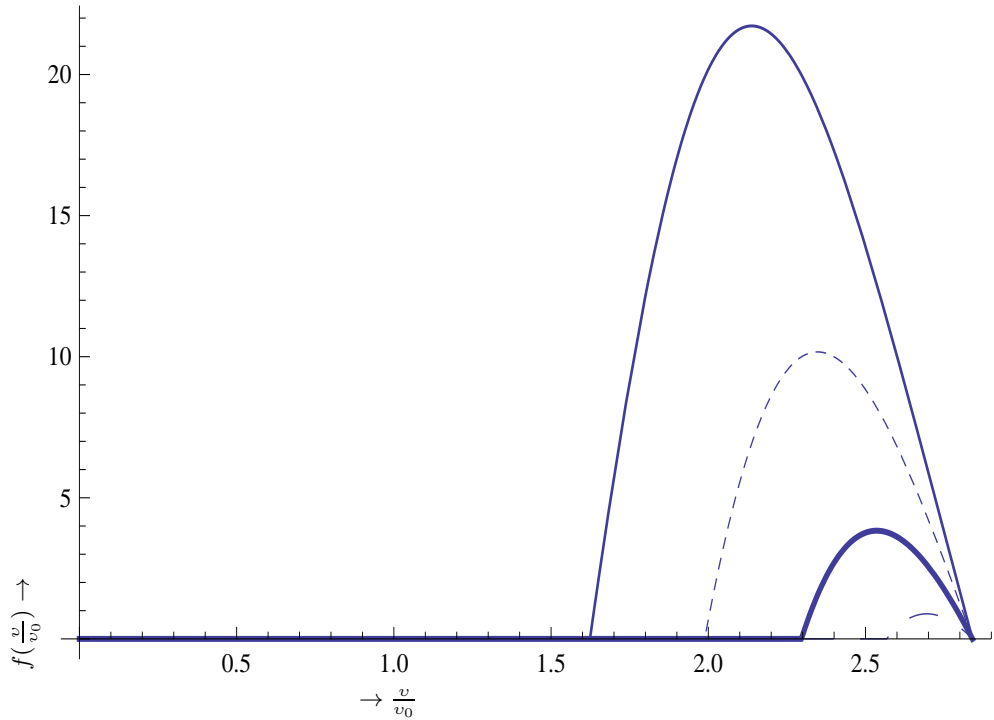}\\
\hspace{-3.0cm}$\rightarrow \frac{\upsilon}{\upsilon_0}$ \\
 \caption{The  velocity distribution obtained with $x=10$. We have selected $T=2$eV  and $b=2,4$ and 6 eV associated with the solid  dashed, and thick solid line respectively. Note the increase in  the scale but the restriction in the velocity range (the $b$=8 eV is barely seen).}
   \label{Fig.vbTx10}
 \end{center}
  \end{figure}

\begin{figure}[!ht]
  \begin{center}
\rotatebox{90}{\hspace{-0.0cm} $f(\frac{\upsilon}{\upsilon_0})\rightarrow$}
\includegraphics[width=0.7\textwidth,height=0.5\textwidth]{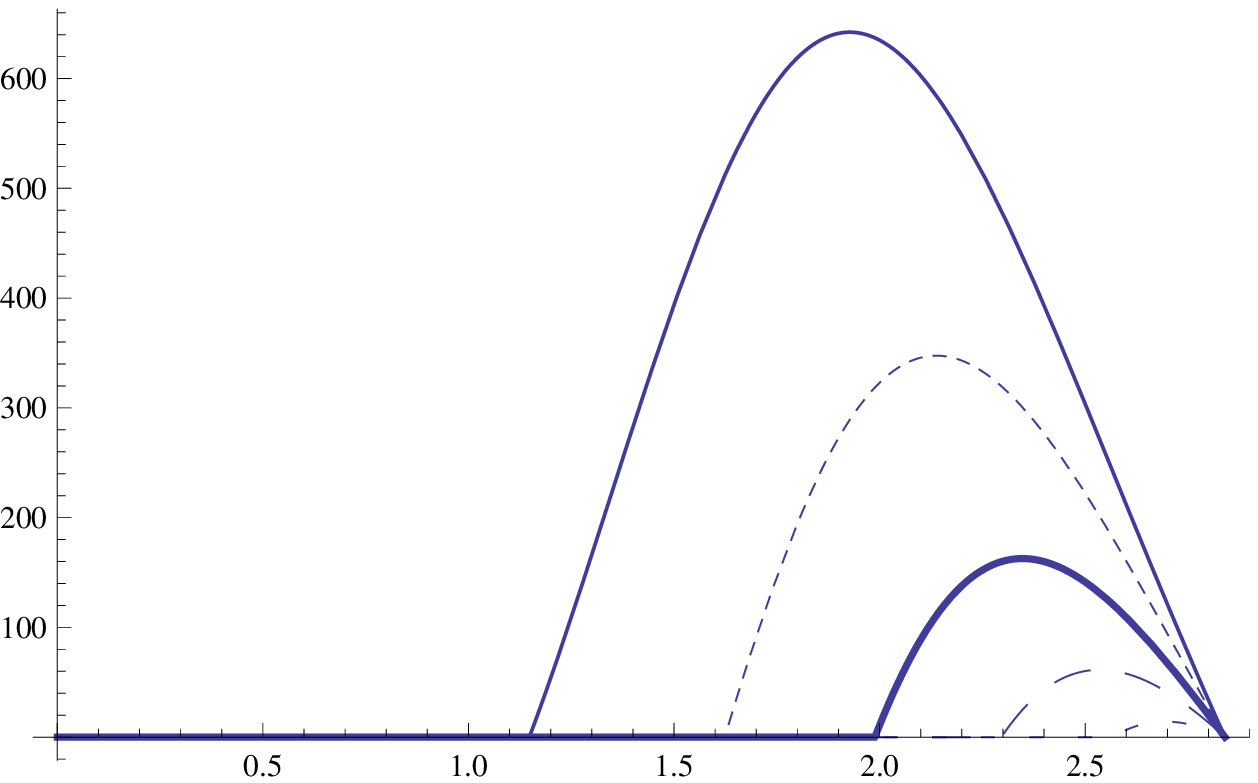}\\
\hspace{-3.0cm}$\rightarrow \frac{\upsilon}{\upsilon_0}$ \\
 \caption{The  velocity distribution obtained with $x=20$. Now we have selected $T=2$eV  and $b=2,6,10,14$ and  18 eV corresponding to the solid,  dashed,  thick solid  and long dashed lines respectively. Note the large increase in  the scale but the restriction in the velocity range (the $b$=18 eV is barely seen).}
   \label{Fig.vbTx20}
 \end{center}
  \end{figure}

This function must be multiplied with the velocity distribution  before proceeding with the needed integrations to obtain the rate. We exhibit the thus resulting   distribution of velocity for various values of $x,b,T$ in Figs \ref{Fig.vbTx2}- \ref{Fig.vbTx20}.


\section{Some results for bound electrons}
\begin{figure}[!ht]
  \begin{center}
\includegraphics[width=0.4\textwidth,height=0.4\textwidth]{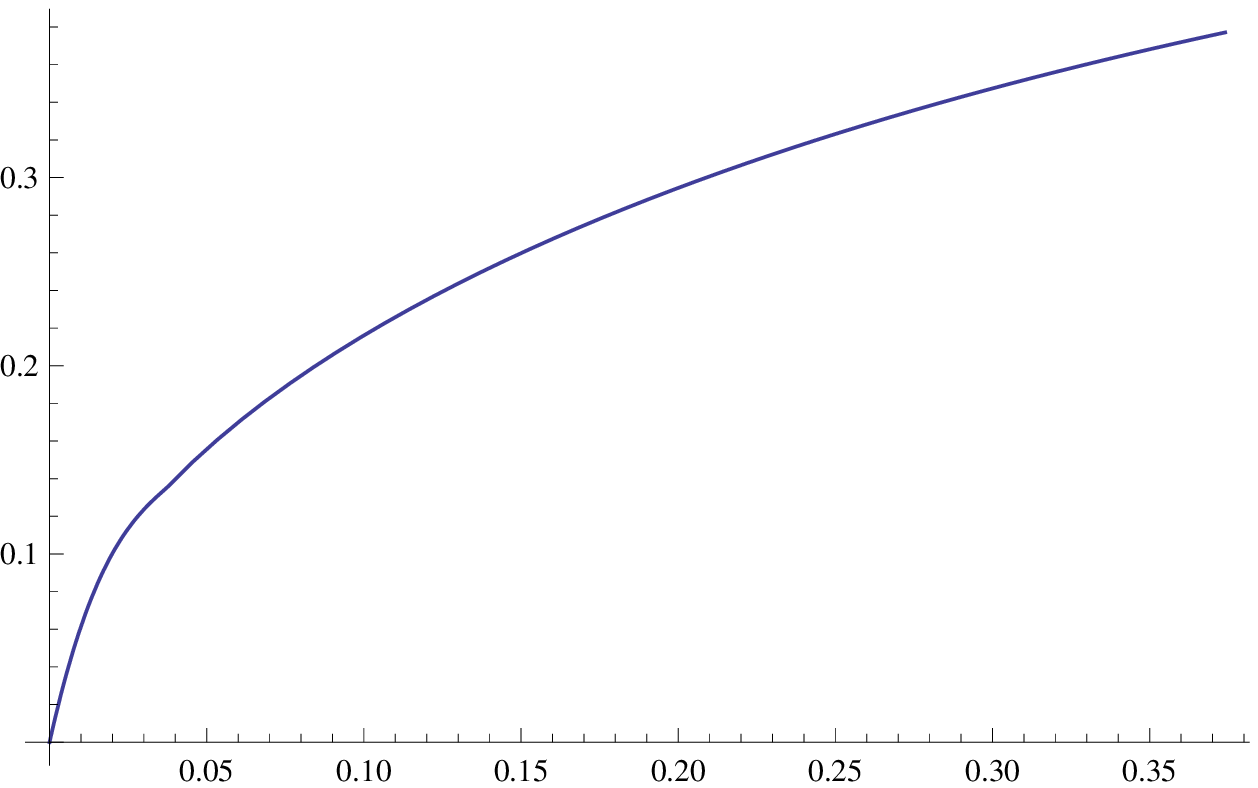}
\\
\hspace{-3.0cm}{$\rightarrow T$ eV }
\caption{The  electron spectrum for bound electrons corresponding $x=1$.
Only electrons with $b<1$ eV can be ejected}. 
   \label{Fig:Sigma0a}
	\end{center}
	\end{figure}

\begin{figure}[!ht]
  \begin{center}
	\subfloat[]
	{
\rotatebox{90}{\hspace{-0.0cm} $\Sigma_0\rightarrow$}
\includegraphics[width=0.4\textwidth,height=0.4\textwidth]{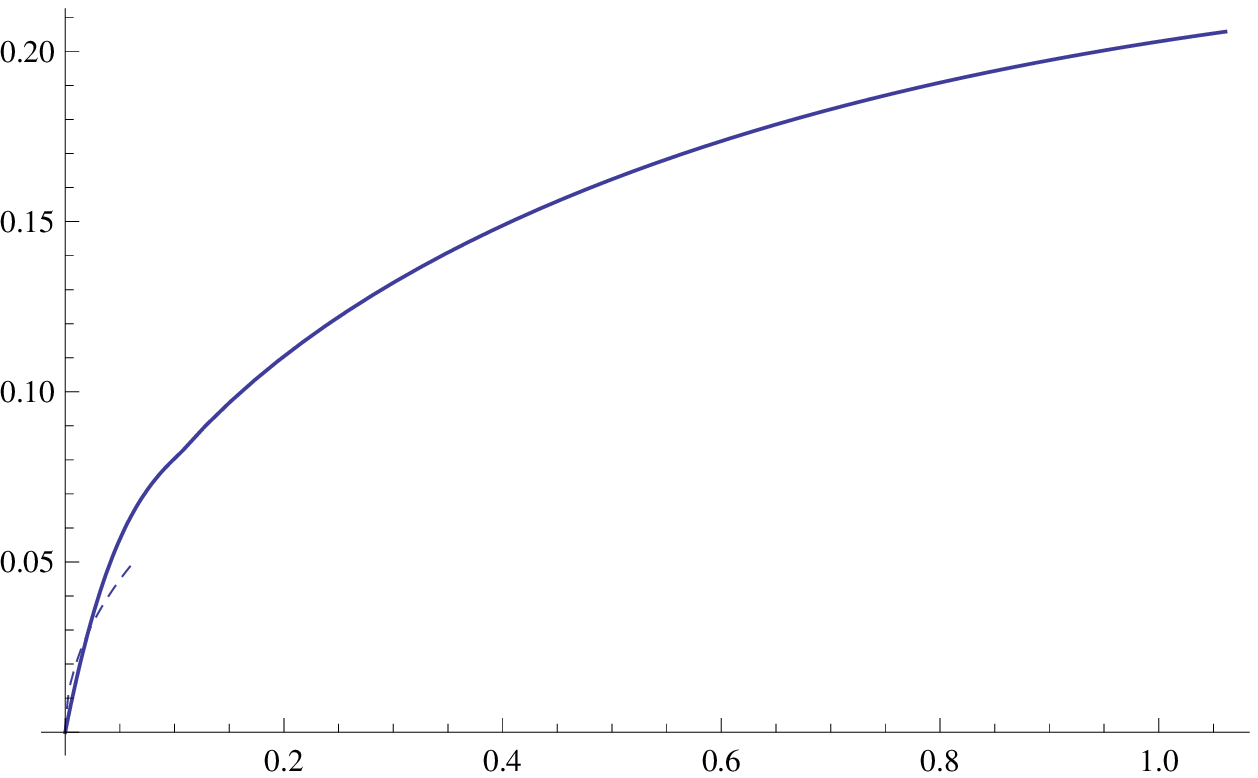}
}
\subfloat[]
	{
\rotatebox{90}{\hspace{-0.0cm} $\Sigma_0\rightarrow$}
\includegraphics[width=0.4\textwidth,height=0.4\textwidth]{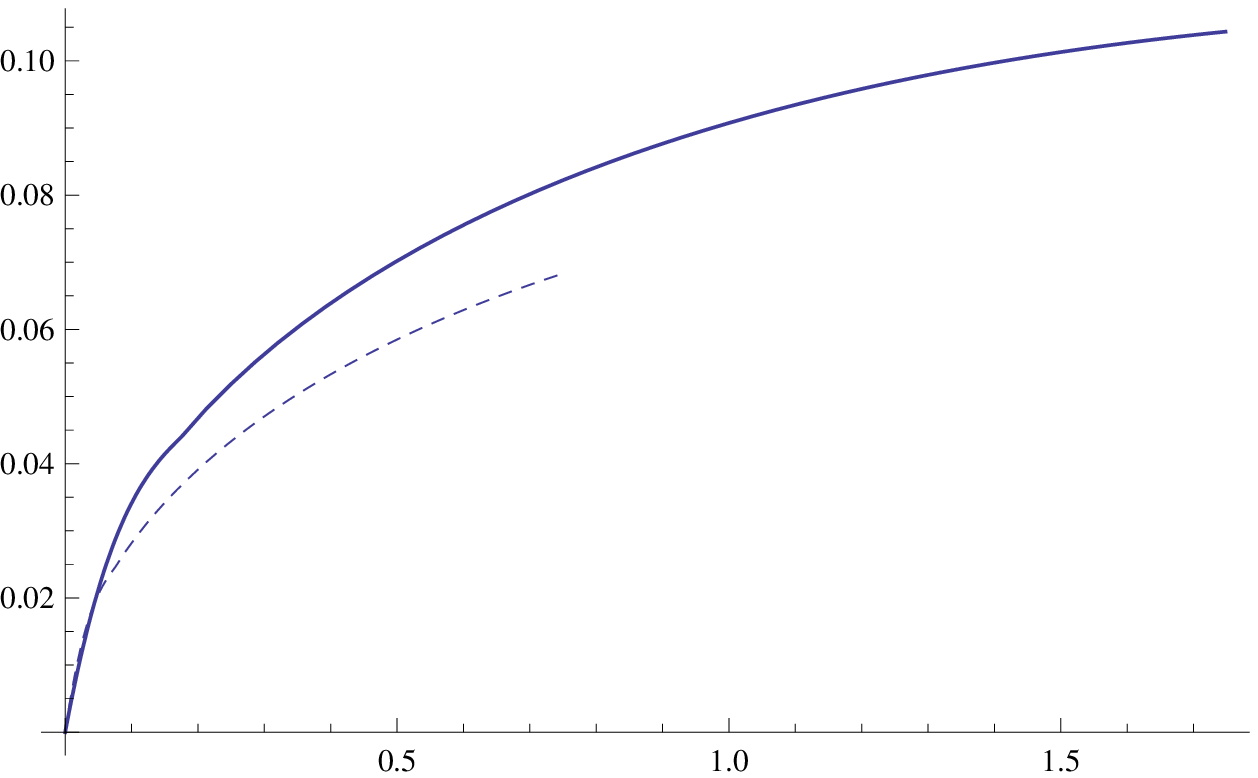}
}\\
\hspace{-3.0cm}{$\rightarrow T$ eV }
 \caption{The  same as in Fig.    \ref{Fig:Sigma0a}, but for $x=10$ in panel (a) and $x=20$ in panel (b). In both cases only electrons with $b<3 $ eV can be ejected. Notice, however, that the spectrum is  suppressed for $b=2$ (dotted line) compared to the corresponding one for $b=1$ (solid line)}. 
\label{Fig:Sigma0b}
 \end{center}
  \end{figure}

\begin{figure}[!ht]
  \begin{center}
	\subfloat[]
	{
\rotatebox{90}{\hspace{-0.0cm} $\Sigma_0\rightarrow$}
\includegraphics[width=0.4\textwidth,height=0.4\textwidth]{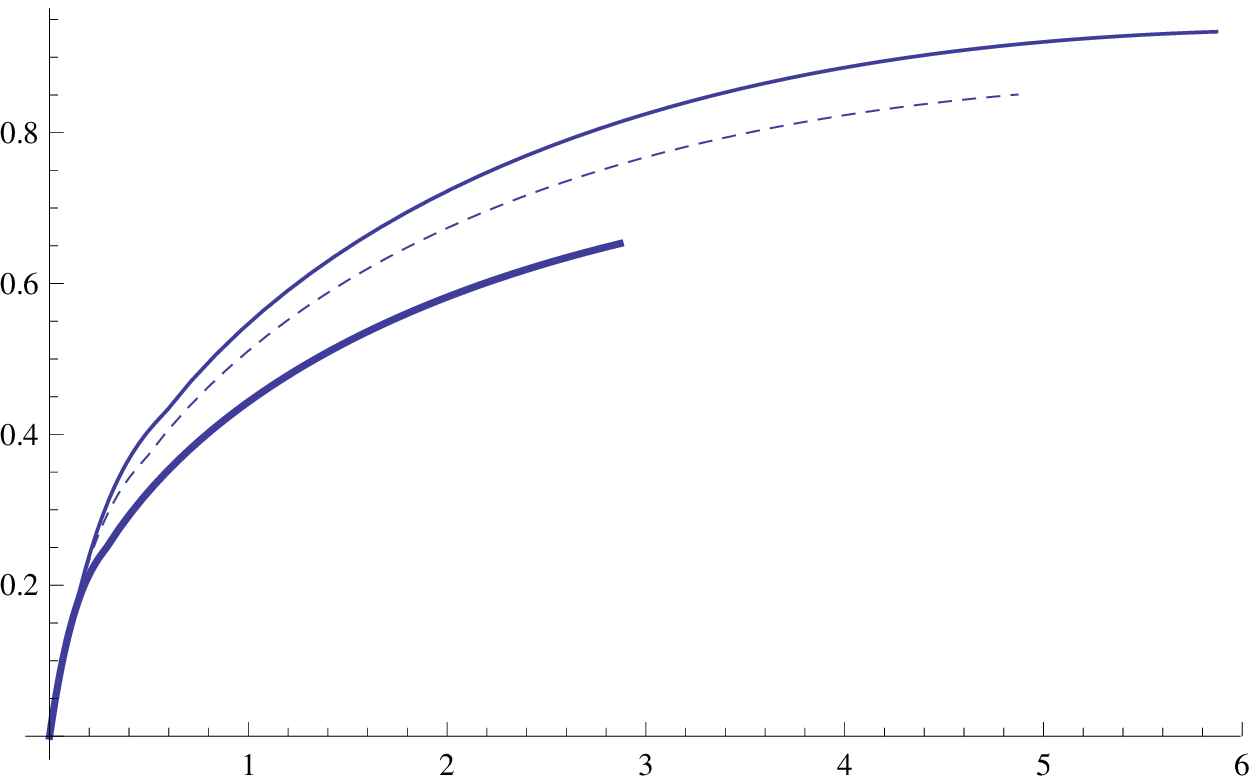}
}
	\subfloat[]
	{
\rotatebox{90}{\hspace{-0.0cm} $\Sigma_0\rightarrow$}
\includegraphics[width=0.4\textwidth,height=0.4\textwidth]{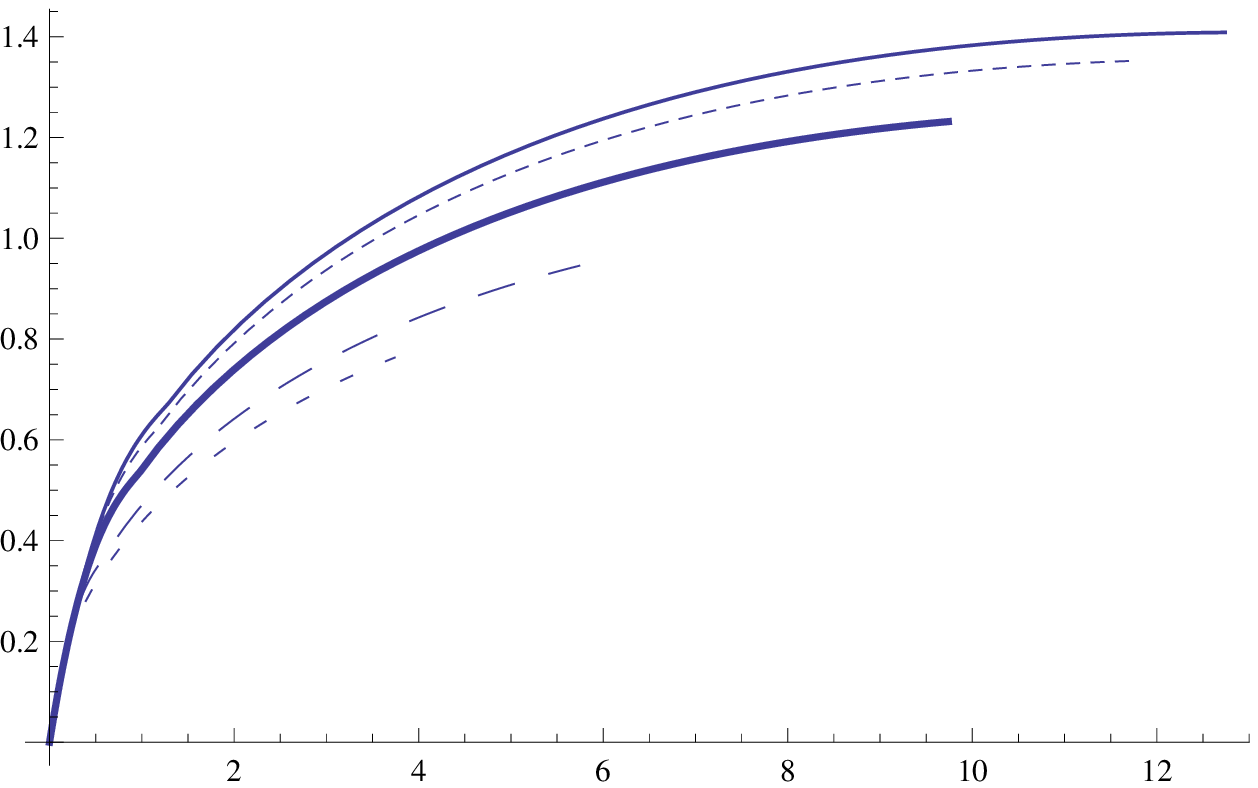}
}\\
\hspace{-3.0cm}{$\rightarrow T$ eV }
 \caption{The  same as in Fig.    \ref{Fig:Sigma0b}, but for $x=50$ in panel (a) and $x=100$ in panel (b). In panel (a) the spectrum is shown for $b=1,\,2$ and 4 eV. In panel (b)  the spectra for $b=1,\,2,\,4$ and 8 eV are shown  with $b$ increasing  downwards. The curve corresponding to $b=10$ is not visible.}
\label{Fig:Sigma0c}
 \end{center}
  \end{figure}

We will limit ourselves to $b\geq 1$ eV and $x\leq  100$
After integrating with the velocity distribution we obtain the electron spectra shown in Figs \ref{Fig:Sigma0a}- \ref{Fig:Sigma0c}

	
\begin{figure}[!ht]
  \begin{center}
\rotatebox{90}{\hspace{-0.0cm} $\frac{\sigma_{av}}{\sigma_0}\rightarrow$}
\includegraphics[width=0.7\textwidth,height=0.5\textwidth]{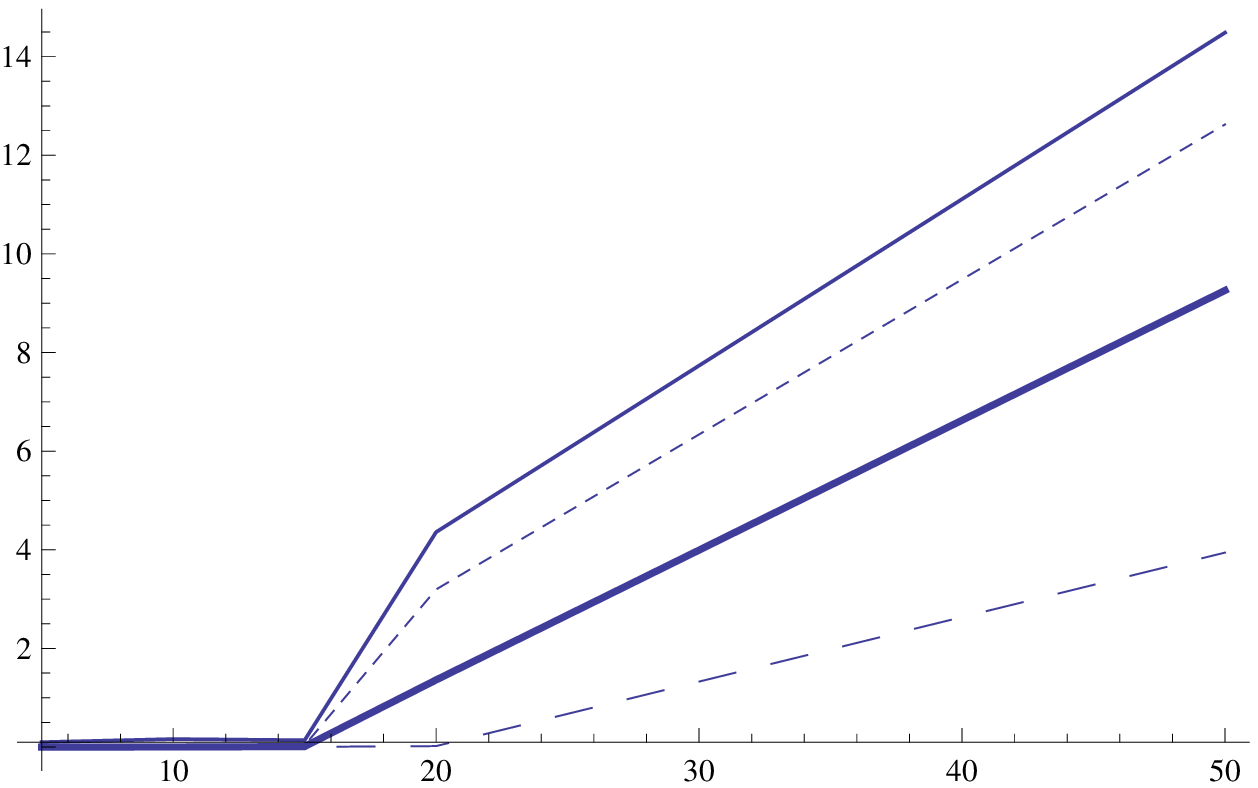}\\
\hspace{-3.0cm}$\rightarrow x=\frac{m_{\chi}}{m_e}$ \\
 \caption{The cross section $\sigma_{av}$ in units of $\sigma_0$ as a function of $x=\frac{m_{\chi}}{m_e}$  for binding energies $b=$1, 2, 4, and 8 eV  increasing  downwards (the curve for $b=10$ is not visible).}  
 \label{Fig.bsigmavsx}
 \end{center}
  \end{figure}

\begin{figure}[!ht]
  \begin{center}
	\subfloat[]
	{
\rotatebox{90}{\hspace{-0.0cm} $\frac{\sigma_{av}}{\sigma_0}\rightarrow$}
\includegraphics[width=0.4\textwidth,height=0.35\textwidth]{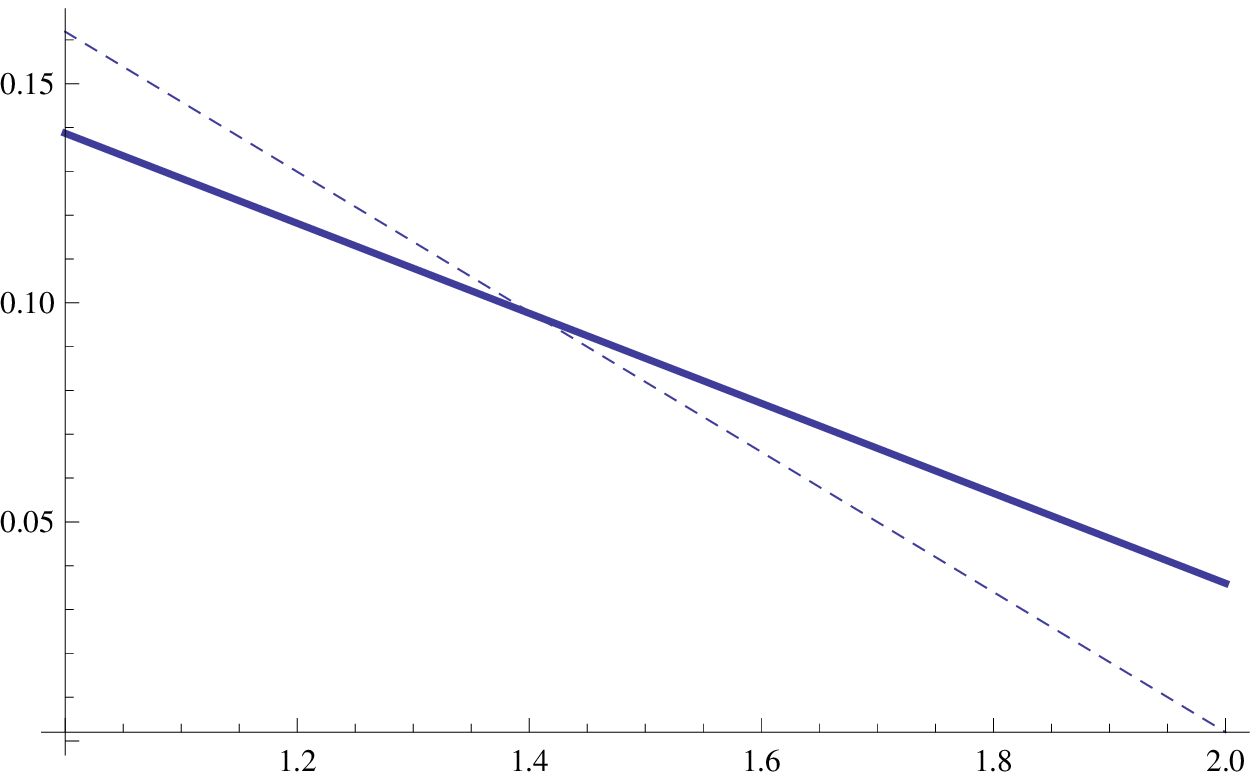}
}
	\subfloat[]
	{
\rotatebox{90}{\hspace{-0.0cm} $\frac{\sigma_{av}}{\sigma_0}\rightarrow$}
\includegraphics[width=0.4\textwidth,height=0.35\textwidth]{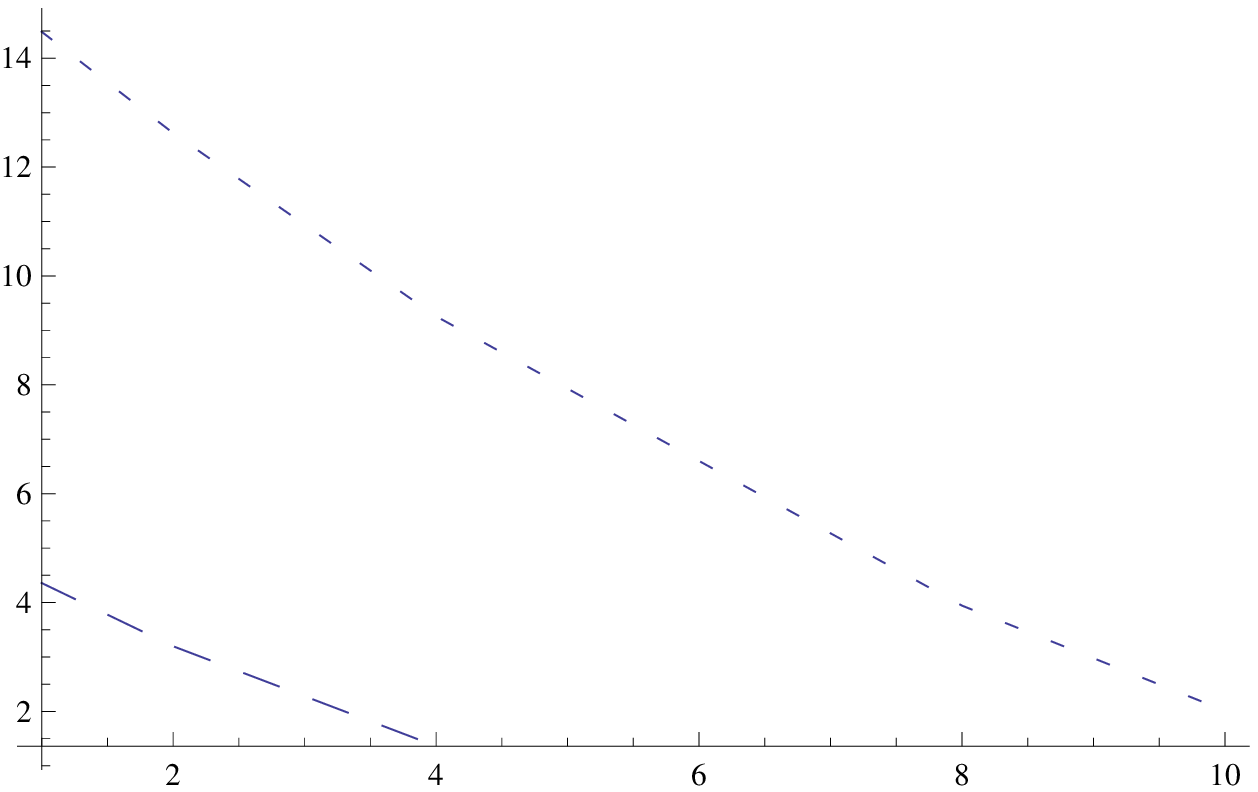}
}
\\
\hspace{-3.0cm}{$\rightarrow b$ eV }
 \caption{The cross section $\sigma_{av}$ in units of $\sigma_0$ as a function of $b$ in eV  for $x=\frac{m_{\chi}}{m_e}=$15 (dotted line) and 20 (solid line) (a) and for $x=\frac{m_{\chi}}{m_e}=$50 and 100 (b) (the lower curve corresponds to the smaller $x$).}
   \label{Fig:xsigmavsb}
 \end{center}
  \end{figure}

	After integrating over the energy spectrum we obtain the cross section $\sigma_{av}$ in units of $\sigma_0$ shown in Fig. \ref{Fig.bsigmavsx} as a function of the WIMP mass for various binding energies. It is perhaps better to show $\sigma_{av}$ as a function of the binding energy.  For $x=10$ only $b\leq 1$ are available. For $b=1$ we find $\sigma_{av}=0.1$ in units of $\sigma_0$. For larger $x$, $\sigma_{av}$ is exhibited in Fig. \ref{Fig:xsigmavsb}.
	
	We thus see that using hydrogenic wave functions we find that, for $m_{\chi}$ greater $10 m_e$, electrons  with $b< 15$ eV become available. The number of electrons with relatively small binding energy for some targets of interest are shown in table \ref{tab:HoELevel}.
\begin{table}
\centering
\caption{The number of electrons with binding energies less than $b_{\rm upb}$ for a given set of targets.}
\label{tab:upb}
\begin{tabular}{|c||c|c|c|c|c|}   \hline \hline
 Target          & $b_{\rm upb}=5$ eV    &  $b_{\rm upb}=10$ eV     & $b_{\rm upb}=15$ eV  &  $b_{\rm upb}=20$ eV   & $b_{\rm upb}=30$ eV  \\ \hline \hline 
$_{9}$F        & -  & -   & -   &	5 & 5 \\
$_{11}$Na    & 1  & 1   & 1   &	1  & 1 \\
$_{32}$Ge    & 4  & 4   & 4   &	4  & 14\\
$_{52}$Te    &  4&   4 & 6   &	6  & 6\\
$_{53}$I       & 5  & 5   & 7   &	7  & 7\\
$_{54}$Xe    &  - &   - & 6   &	6  & 8 \\
$_{83}$Bi    &  3 &   5 & 5   & 5	 &  15 \\
  \hline \hline
\end{tabular}
\label{tab:HoELevel}        
\end{table}
	Anyway there seem to be  
	which leads    $Z_{eff}=5$ for a suitable  atom with large $Z$.  Thus for $Z_{eff}=5$  one can conservatively  $\sigma_{av}$ to be $\approx 0.15$, which leads to about 1 event per kg-y compared to the 3 per kg-y we got above for lighter WIMPs without the electron binding.
	The Fermi function is incorporated into the results.
	
\section{discussion}
	\label{sec:discussion}
	We have seen that the use of electron detectors may be a good way to directly detect light WIMPs in the MeV region. The electron density in our vicinity is very high, the elementary WIMP-electron cross section section may be quite large and the event rate may be further enhanced  by the behavior of the Fermi function at low energies. With bound atomic electrons, however, there seems to be a problem, because a small fraction of electrons,  $Z_{eff}/Z$, can be exploited,  with $Z_{eff}$  being those electrons with binding energies  below the 15 eV range. This is  reminiscent of the difficulty encountered in the inelastic WIMP nucleus scattering, whereby only very low excited states can be reached.
	We have seen that the expected rates for low energy electron recoils due to light WIMPs are sensitive to the Fermi function corrections. In fact the inclusion of such corrections may increase the rate by factors of 8 for low energy electrons. So event rates of about 1 to 3 events per kg-y are possible.
	
It has recently been suggested that it  is possible to detect even very light WIMPS, much lighter than the electron, utilizing Fermi-degenerate materials like superconductors\cite{HPZ15}. In this case the energy required is essentially the gap energy of about $1.5 kT_c$, which is in the meV region, i.e the electrons are essentially free. The authors are perhaps aware of the fact that not all the  kinetic  energy of the WIMP can be transferred to the system. As we have seen the maximum fraction is approximately 1/3 and occurs  if the mass of the WIMP is equal to $m_e$,  see Eq.~(\ref{Eq:Ttransfratio}) for $x=1$. These authors  probably have a way to circumvent the fact that  a small amount  of energy  will be deposited, partly because a  small fraction of energy of the WIMP  will be transferred  to their system (see Fig. \ref{fig:Txtometransf}) and also because the average energy of the WIMP is smaller. 
Anyway, if they manage to accumulate a large number of electrons  in their targets, the obtained rates maybe sufficient. 
More recently it is claimed that even smaller energies in meV can be detected in the case of 
Liquid Helium~\cite{SchZur16}. The expected event rates and the total energy deposited in such essentially bolometer type detectors are currently being estimated more precisely  and they will appear elsewhere.


It thus appears that light WIMPs in the MeV region can, in principle,  be detected. The detection techniques and targets employed, however,  may have to be different than the  ones employed in standard WIMP searches.

\section*{Acknowledgments} 
J.D.V  is happy to acknowledge support of this work by 
i)  the National Experts Council of China via  a "Foreign Master" grant and 
ii)  IBS-R017-D1-2016-a00 in the Republic of Korea. 
A substantial part of this work  performed while J.D.V. was on a visit to the University of Adelaide, supported by CoEPP and the Centre for the Subatomic Structure of Matter (CSSM). He is  happy to thank   Professors Yannis K. Semertzidis  of KAIST, Anthony Thomas of Adelaide and Edna Cheung of Nanjing University for their hospitality. \\
Y.-K.E.C acknowledges support from  the Jiangsu Ministry of Science and Technology under contract BK20131264, 
and the Priority Academic Program Development for Jiangsu Higher Education Institutions (PAPD).

\end{document}